\documentclass[twocolumn,twocolappendix]{aastex631}
\usepackage{CJK}

\usepackage{amsmath}
\usepackage{appendix}
\received{October 10, 2023}
\revised{December 21, 2023}
\accepted{December 22, 2023}

\DeclareGraphicsExtensions{.eps,.eps.gz,.epsi}
\newcommand{\teff}{\mbox{$T_{\rm eff}$}}
\newcommand{\logg}{{\rm{log}~$g$}}
\newcommand{\logginequ}{{\rm{log}~g}}
\newcommand{\feh}{{\rm [Fe/H]}} 
\newcommand{\ebv}{$E(B-V)$} 
\newcommand{\rv}{$R(V)$}

\shorttitle{A comprehensive correction of the {\it Gaia} DR3 XP spectra}
\shortauthors{Huang et al.}

\graphicspath{{./}{my_figures/}}

\begin{document}
\begin{CJK*}{UTF8}{gbsn}
\title{A comprehensive correction of the {\it Gaia} DR3 XP spectra}

\author[0000-0002-1259-0517]{Bowen Huang (黄博闻)}
\affiliation{Institute for Frontiers in Astronomy and Astrophysics, Beijing Normal University, Beijing, 102206, China}
\affiliation{Department of Astronomy, Beijing Normal University No.19, Xinjiekouwai St, Haidian District, Beijing, 100875, China}

\author[0000-0003-2471-2363]{Haibo Yuan (苑海波)}
\affiliation{Institute for Frontiers in Astronomy and Astrophysics, Beijing Normal University, Beijing, 102206, China}
\affiliation{Department of Astronomy, Beijing Normal University 
No.19, Xinjiekouwai St,
Haidian District, Beijing, 100875, China}

\correspondingauthor{Yuan Haibo (苑海波)}
\email{yuanhb@bnu.edu.cn}

\author[0000-0002-5818-8769]{Maosheng Xiang (向茂盛)}
\affiliation{National Astronomical Observatories, Chinese Academy of Sciences, 20A Datun Road, Chaoyang District, Beijing, China}
\author[0000-0003-3250-2876]{Yang Huang (黄样)}
\affiliation{School of Astronomy and Space Science, University of Chinese Academy of Sciences, Beijing, 100049, China}

\author[0000-0001-8424-1079]{Kai Xiao (肖凯)}
\affiliation{Institute for Frontiers in Astronomy and Astrophysics, Beijing Normal University, Beijing, 102206, China}
\affiliation{Department of Astronomy, Beijing Normal University No.19, Xinjiekouwai St, Haidian District, Beijing, 100875, China}
\affiliation{School of Astronomy and Space Science, University of Chinese Academy of Sciences, Beijing, 100049, China}
\author[0000-0003-3535-504X]{Shuai Xu (徐帅)}
\affiliation{Institute for Frontiers in Astronomy and Astrophysics, Beijing Normal University, Beijing, 102206, China}
\affiliation{Department of Astronomy, Beijing Normal University No.19, Xinjiekouwai St, Haidian District, Beijing, 100875, China}
\author[0000-0003-1863-1268]{Ruoyi Zhang (张若羿)}
\affiliation{Institute for Frontiers in Astronomy and Astrophysics, Beijing Normal University, Beijing, 102206, China}
\affiliation{Department of Astronomy, Beijing Normal University No.19, Xinjiekouwai St, Haidian District, Beijing, 100875, China}
\author[0000-0002-9824-0461]{Lin Yang (杨琳)}
\affiliation{Department of Cyber Security, Beijing Electronic Science and Technology Institute, Beijing, 100070, China}
\author[0000-0002-3651-0681]{Zexi Niu (牛泽茜)}
\affiliation{School of Astronomy and Space Science, University of Chinese Academy of Sciences, Beijing, 100049, China}
\affiliation{National Astronomical Observatories, Chinese Academy of Sciences, 20A Datun Road, Chaoyang District, Beijing, China}
\author[0009-0007-5610-6495]{Hongrui Gu (顾弘睿)}
\affiliation{CAS Key Laboratory of Optical Astronomy, National Astronomical Observatories, Chinese Academy of Sciences, Beijing 100101, China}
\affiliation{School of Astronomy and Space Science, University of Chinese Academy of Sciences, Beijing, 100049, China}

\begin{abstract} 
By combining spectra from the CALSPEC and NGSL, as well as spectroscopic data from the LAMOST Data Release 7 (DR7), we have analyzed and corrected the systematic errors of the Gaia DR3 BP/RP (XP) spectra. The errors depend on the normalized spectral energy distribution (simplified by two independent ``colors'') and $G$ magnitude. Our corrections are applicable in the range of approximately $-0.5<BP-RP<2$, $3<G<17.5$ and $E(B-V)<0.8$. 
To validate our correction, we conduct independent tests by comparing with the MILES and LEMONY spectra. The results demonstrate that the systematic errors of $BP-RP$ and $G$ have been effectively corrected, especially in the near ultraviolet. The consistency between the corrected Gaia XP spectra and the MILES and LEMONY is better than 2 per cent in the wavelength range of $336-400$\,nm and 1 per cent in redder wavelengths.
A global absolute calibration is also carried out 
by comparing the synthetic Gaia photometry from the corrected XP spectra with the corrected Gaia DR3 photometry. 
Our study opens up new possibilities for using XP spectra in many fields. 
A Python package is publicly available to do the corrections (https://doi.org/10.12149/101375 or https://github.com/HiromonGON/GaiaXPcorrection).

\end{abstract}

\keywords{catalogs — instrumentation: spectroscopy --  extinction -- methods: data analysis -- surveys -- techniques: imaging, spectroscopic}

\section{Introduction} 

Gaia Data Release 3 (DR3; \citealt{dr3content}) has made an extensive and novel data-set available, including about 220 million flux-calibrated low-resolution BP/RP (XP; Blue Photometer and Red Photometer) spectra that cover the wavelength range of $336-1020$\,nm with a measured spectral resolution of $R \sim 20-70$. It provides one of the best data-sets and a unique opportunity to measure stellar parameters of enormous number of stars, to probe interstellar extinction and investigate Galactic structure and formation history in great detail, and to perform high-precision photometric calibration of other surveys.

Despite their huge potential, Gaia XP spectra  have been found to suffer systematic errors that depend on color and $G$ magnitude (\citealt{dr3extcali}). 
For instance, color-dependent errors between red and blue sources have been observed to exceed 50 per cent in the $336-400$\,nm range. Systematic errors, characterized by a discontinuity, have also been noticed in the wavelength range of $640-680$\,nm, where the imperfect conjunction between BP and RP spectra can be observed.
These errors have limited the optimal usage of the XP spectra in various areas, such as the determination of stellar parameters (\citealt{GaiaPara}), extinction curves, and synthetic magnitudes (\citealt{GSPC}).

Therefore, in this work, we aim to analyze and correct the systematic errors of the Gaia DR3 XP spectra. 
To achieve this goal, we compared the XP spectra with external spectral libraries, including the CALSPEC (\citealt{CALSPEC14}; \citealt{CALSPEC22}) and Hubble's Next Generation Spectral Library(NGSL ;  \citealt{NGSL2007}).
We also used spectroscopic data from the LAMOST  DR7 (\citealt{LAMOST}) to supplement and extend our correction to sources of $14 < G < 17.5$. 
We have found that the systematic errors depend on the normalized spectral energy distribution (SED) of Gaia DR3 XP spectra (described by two independent ``colors'') and $G$ magnitude.
We also conducted independent tests by comparing 
the corrected Gaia XP spectra with the Medium-resolution Isaac Newton Telescope library of empirical spectra (MILES; \citealt{MILES}) and  a library of empirical medium-resolution spectra by observations with the NAOC Xinglong 2.16-m and YNAO Gaomeigu 2.4-m telescopes (LEMONY; \citealt{LEMONY}).

This paper is organized as follows.
In Section\,2, we introduce the spectroscopic datasets used in our study.
In Section\,3, we describe how we derived the systematic errors using CALSPEC and NGSL.
In Section\,4, we extend our correction to sources faint to $G = 17.5$ using spectroscopic data from LAMOST DR7.
In Section\,5, we provide a comprehensive analysis of the systematic errors in the XP spectra and perform independent tests on the corrected Gaia XP spectra using MILES and LEMONY.
In Section\,6, we carry out a global absolute calibration by comparing the corrected synthetic photometry and the Gaia DR3 photometry.
The conclusions are given in Section\,7.

\section{Data Sets} 

\subsection{ {\it Gaia} DR3 and XP spectra} 
{\it Gaia} is a satellite  mission of the European Space Agency. It focuses on astrometry, photometry, and spectroscopy of objects in the Milky Way and Local Group (\citealt{GaiaMission2016a}). The Gaia DR3 provides approximately 220 million XP spectra covering the wavelength range of $336-1020$\,nm (\citealt{GaiaDR32022}; \citealt{GaiaXPVad}; \citealt{dr3intcali}).

Gaia XP spectra take the form of a projection onto 55 orthonormal Hermite functions as base functions for both the BP and RP spectra, rather than the flux as a function of wavelength (\citealt{dr3intcali}; \citealt{dr3extcali}). Therefore, each XP spectrum consists of 110 coefficients. For our work, we transform the coefficients into wavelength space to enable comparison with external spectra libraries. Therefore, we use the \texttt{GaiaXPy} package (\citealt{GaiaXPy}) to transform the coefficients into default sampled spectra, in wavelength space ranging from $336-1020$\,nm in increments of $2 $\,nm.
In this transformation, the package offers a truncation option to remove spurious features in the spectra, particularly for faint sources. It is because that truncated coefficients of higher-order bases are consistent with noise (\citealt{{GaiaXPVad}}). To facilitate the use of our correction, we have considered both nontruncated and truncated forms of XP spectra in the subsequent steps. 

\subsection{CALSPEC} 
CALSPEC (\citealt{CALSPEC14}) contains 92 composite stellar spectra that are flux standards on the Hubble Space Telescope (HST) system, including main sequence stars, subdwarfs and white dwarfs. The dataset utilized in our work corresponds to the latest 2023 September Update version of CALSPEC.
The colors and magnitudes of these sources, expressed in terms of Gaia's $BP-RP$ and $G$, range roughly from $-0.5<BP-RP<5$ and $2<G<17$, with the majority of sources concentrated in the range of $-0.5<BP-RP<1$ and $4<G<14$. The most precise and internally consistent flux measurements for CALSPEC in the near-ultraviolet to the near-infrared are obtained from HST spectrophotometry, mainly from the Space Telescope Imaging Spectrograph (STIS) with a resolution of $R \sim 800$. The relative fluxes of primary reference standards are accurate to within 1 per cent. This set of spectra is utilized as flux standards in our work for correcting the Gaia XP spectra.

\subsection{NGSL} 
NGSL (\citealt{NGSL2007}) comprises STIS spectra of 378 stars, covering the wavelength range of $180-1010$\,nm with a resolution of $R\sim 1000$. Similarly, the colors and magnitudes of these sources, expressed in terms of Gaia's $BP-RP$ and $G$, range roughly from $-0.5<BP-RP<2$ and $3<G<10$.  The majority of the NGSL spectra have been well-calibrated, achieving a precision of 2--3 per cent in their absolute fluxes (\citealt{NGSL2009}). The spectra from NGSL are combined with those from CALSPEC for correcting the Gaia XP spectra.

\subsection{LAMOST DR7} 
LAMOST is a 4 m quasi-meridian reflective Schmidt telescope with 4000 fibers, operated by the National Astronomical Observatory of China, Chinese Academy of Sciences. A total number of 10,640,255 low-resolution spectra covering the whole optical wavelength range of $369-910$\,nm at a spectral resolution of $R \sim 1800$ is contained in the LAMOST DR7 $\footnote{http://dr7.lamost.org/v1.3/}$ . The basic stellar parameters such as effective temperature ${\teff}$, surface gravity \logg, metallicity ${\feh}$ for late A and FGK stars are derived by the LAMOST Stellar Parameter Pipeline (\citealt{LAMOST}, \citealt{LASP}). The internal errors of LAMOST parameters at signal-to-noise ratios greater than 20 are
${\teff} \sim 50-100\,K$, \logg, $\feh \sim 0.05-0.1 \,dex$ 
(\citealt{Niu2021a}). The colors and magnitudes of these sources range roughly from $0<BP-RP<3$ and $9<G<17.8$. 
To overcome limitations caused by insufficient numbers of faint sources from CALSPEC and NGSL, we use spectroscopic information from LAMOST DR7 as a supplement.

\subsection{MILES}
The MILES dataset consists of 985 spectra acquired from the 2.5 m Isaac Newton Telescope, covering the range of $352.5-750$\,nm with a spectral resolution of $0.25 $\,nm
(\citealt{MILES2011}).
The colors and magnitudes of MILES sources are roughly in the range of $-0.5<BP-RP<5$ and $2<G<18$, with the majority of sources concentrated in the range of $-0.5<BP-RP<3$ and $2<G<14$.
The spectra from MILES are utilized to verify the accuracy of our XP spectra correction.

\subsection{LEMONY}
The LEMONY (\citealt{LEMONY}) dataset were collected using two long-slit spectrographs covering the wavelength range of $380-900$\,nm. The first spectrograph, OMR (Manufactured by Optomechanics Research, Inc), was mounted on the $2.16$\,m telescope of the National Astronomy Observatory of China, covering $380-518$\,nm; while the second one, YFOSC, was mounted on the $2.4$\,m telescope of the Yunnan Astronomical Observatory, covering $515-900$\,nm. The combined dataset contains 822 spectra from OMR and 1324 spectra from YFOSC, with an average spectral resolution of $0.33 $\,nm and a mean signal-to-noise ratio per pixel exceeding 100. A total of 629 spectra were merged from the two spectrographs.
The colors and magnitudes of LEMONY sources are roughly in the range of $0<BP-RP<3$ and $4<G<10$, with the majority of sources concentrated in the range of $0<BP-RP<1.5$ and $4<G<10$.
The spectra from LEMONY are also utilized to verify the accuracy of our XP spectra correction.

\section{Correction by CALSPEC and NGSL} 
\subsection{Data Selection}
\begin{figure}[ht]
\centering
\includegraphics[width=80mm]{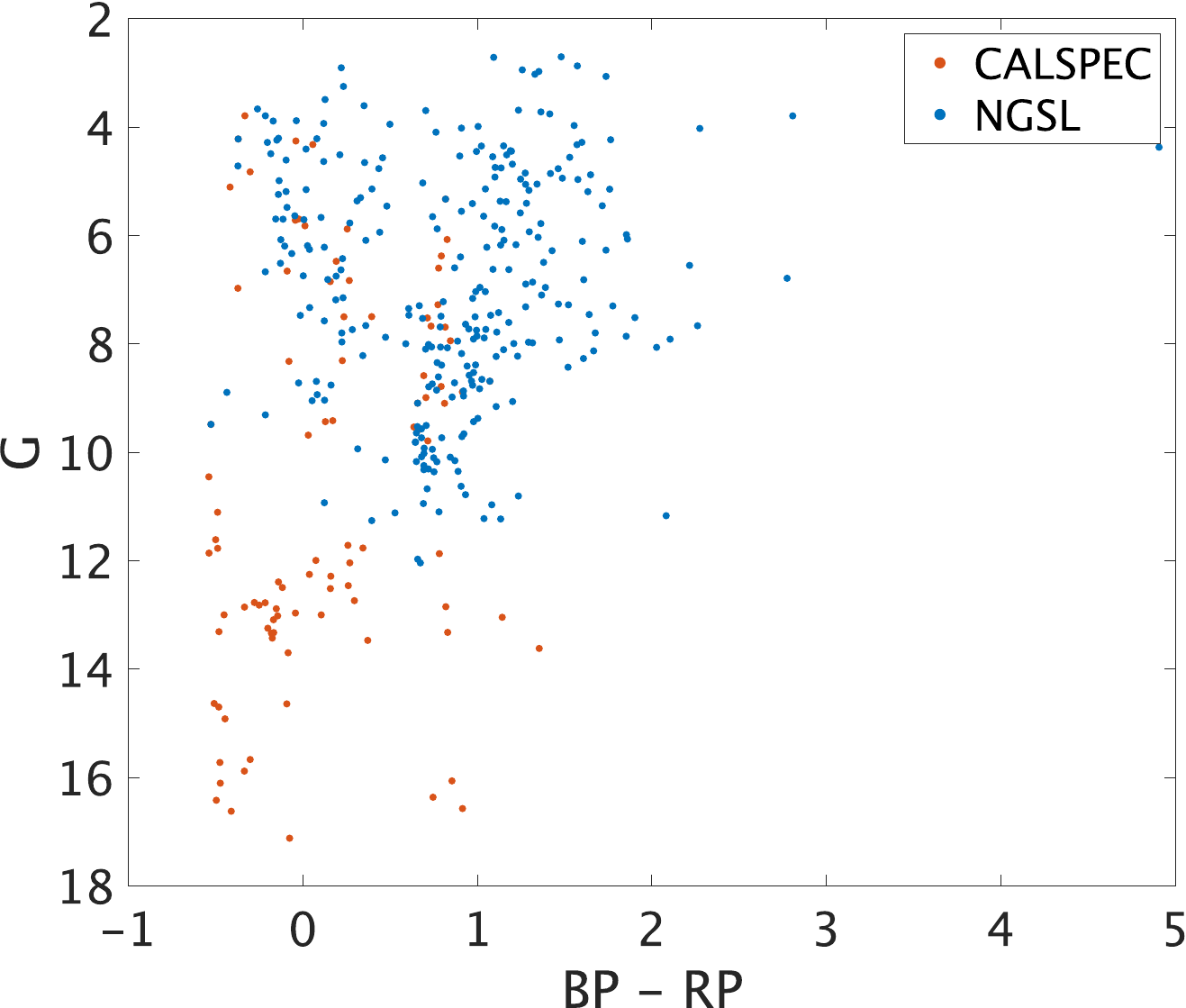}
\caption{Color-magnitude diagram of the training sample. Red and blue points represent sources from the CALSPEC and NGSL, respectively.} 
\label{Fig1}
\end{figure}

We cross-match CALSPEC and NGSL with Gaia DR3 XP spectra by searching for the closest star to a maximum angular separation of 5 arcsec. 
A total number of 94 CALSPEC sources and 276 NGSL sources are selected as the training (reference) sample. Their properties are presented in Figure \ref{Fig1}. Most sources in the training sample have $E(B-V)<0.8$. The \ebv~values of the CALSPEC sources are from the dust reddening map of Schlegel et al. (SFD98; \citealt{SFD98}), while those for NGSL sources are from \cite{NGSL}.

\subsection{Data Processing}
As the spectral resolution of the training sample differs from that of the Gaia XP spectra, it is necessary to convolve the spectra of the training sample to match the Gaia XP spectra.
The wavelength of the training sample covers the blue end of the XP spectra well, but not the red end for the NGSL sources and a few CALSPEC sources. 
To mitigate the border effect of convolution in the subsequent process, we extrapolate the flux of training sample spectra to $1200 $\,nm when necessary, and artificially assign the extrapolated values to be the flux value of the reddest pixel.
Subsequently, we perform a convolution of the CALSPEC and NGSL spectra, which have respective resolutions of $R = 800$ and $R = 1000$ to match the resolution of the XP spectra (refer to Figure 18 and Table 1 in \citealt{dr3extcali}). We then interpolate the training sample onto the sampled wavelength of the XP spectra using a cubic spline.

\subsection{Method and result}
\begin{equation} 
Systematic \,\, Error = \frac{Flux_{\,CALSPEC \,,\,NGSL}}{Flux_{\,Gaia \,XP}}
\label{SE1}
\end{equation}
We assume that the discrepancies between the training sample and their corresponding XP spectra are attributed to the systematic errors in the latter, as shown in Equation \ref{SE1} for each wavelength. We establish the relations between the systematic errors and the source properties in the subsequent steps.

\begin{figure*}[ht]
\centering
\includegraphics[width=160mm]{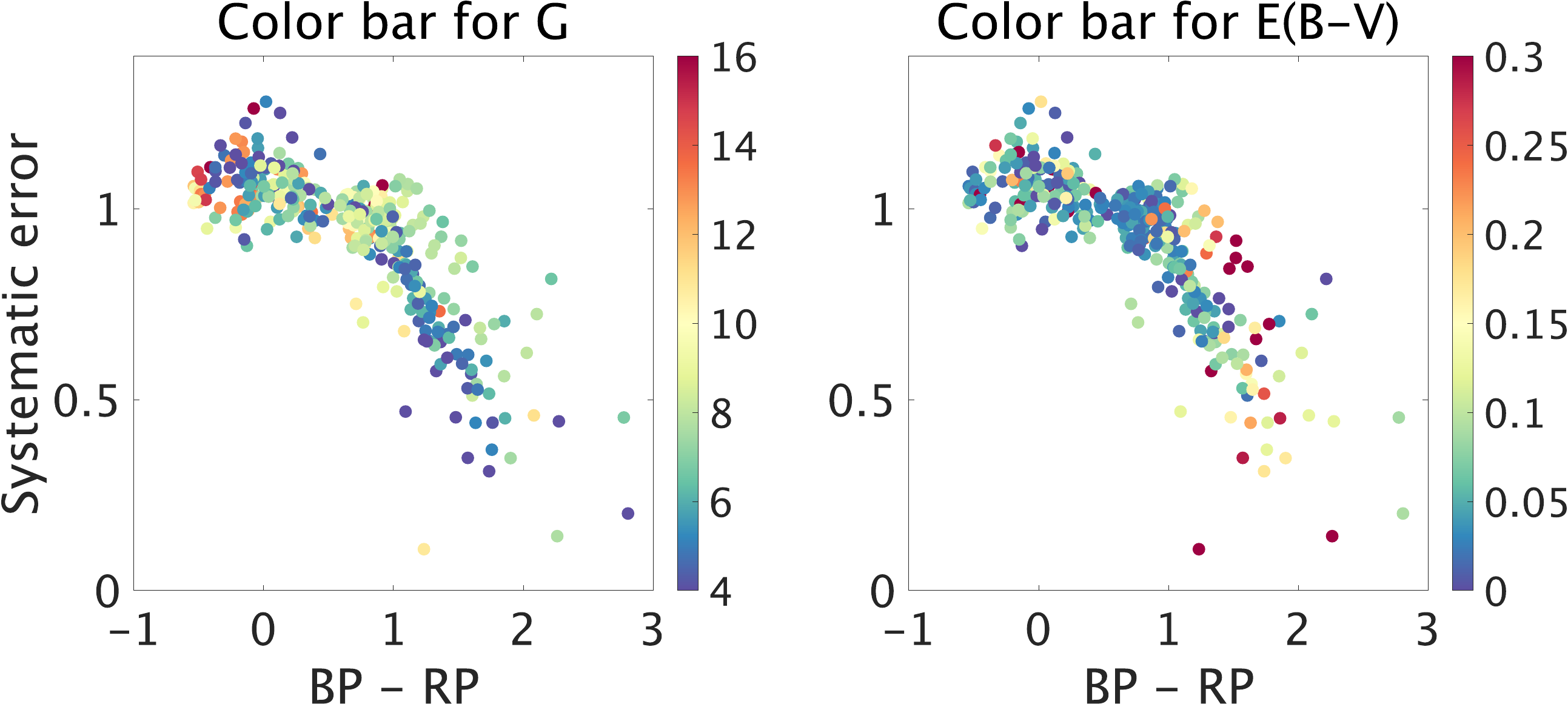}
\caption{Systematic errors at $360$\,nm as a function of $BP-RP$. The truncated XP spectra are adopted here. 
In the left panel, the color of the points denotes the $G$ magnitude, whereas in the right panel, it stands for the values of \ebv.}
\label{Fig2}
\end{figure*}

\begin{figure*}[ht]
\centering
\includegraphics[width=160mm]{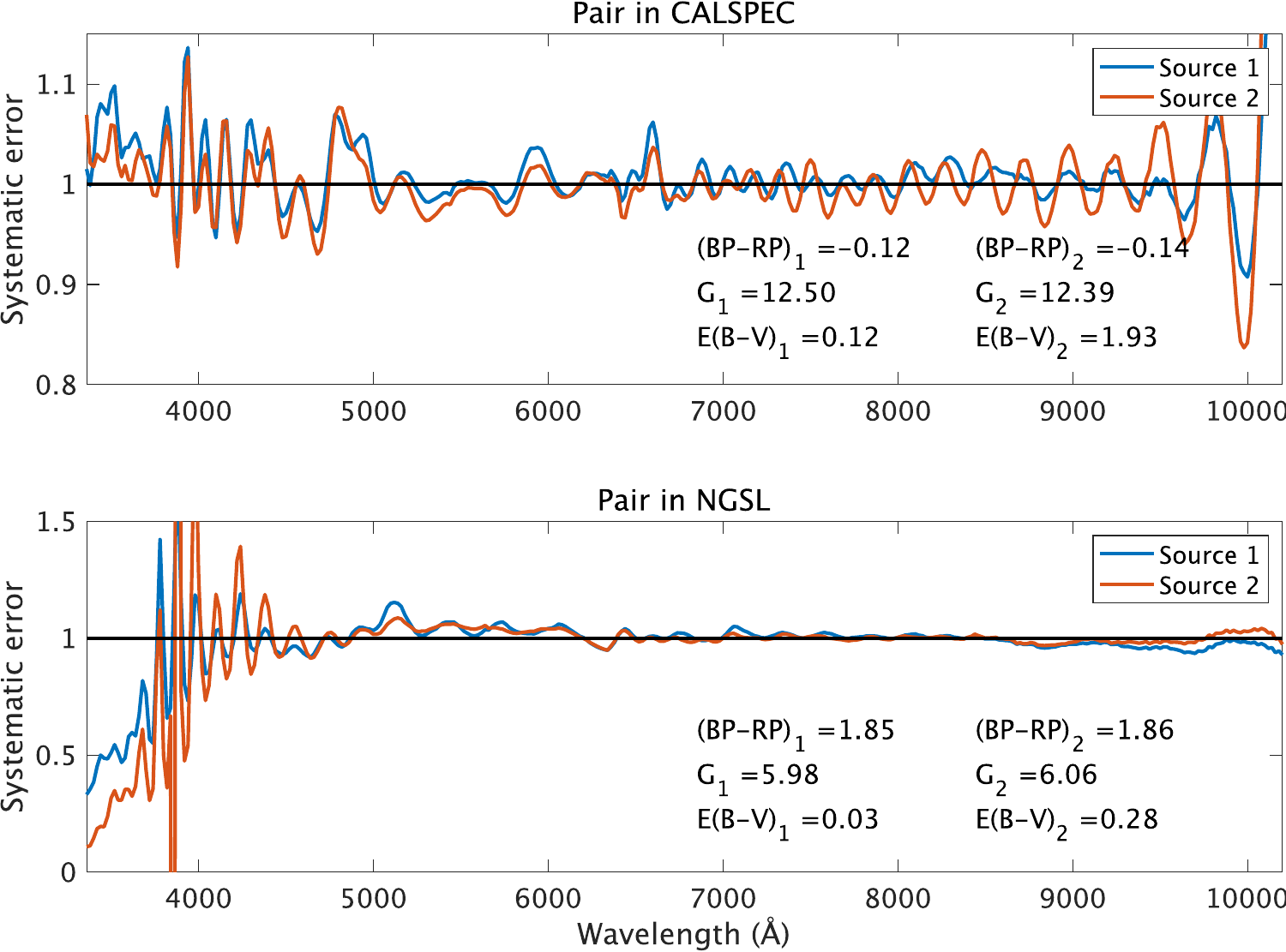}
\caption{Comparison of systematic errors between two pairs of sources from CALSPEC (top panel) and NGSL (bottom panel). The sources in each pair have similar $BP-RP$ color and $G$ magnitude, but very 
different \ebv.  The low-extinction and high-extinction sources are represented by blue and red lines, respectively.} 
\label{Fig3}
\end{figure*}

As already demonstrated in \citealt{dr3extcali}),  the systematic errors depend on color $BP - RP$ as well as $G$ magnitude. However, we find that the systematic errors also depend on the reddening $E(B-V)$.
As illustrated in Figure \ref{Fig2}, the systematic errors show a bifurcation at $1<BP-RP<2$, which is attributed to different reddening values. 
Figure \ref{Fig3} compares two pairs of sources that have similar $BP-RP$ and $G$ values, but differ significantly in \ebv ($\Delta BP-RP < 0.05$, $\Delta G < 0.15$, and $\Delta$ \ebv $> 0.2$). One pair is selected from the CALSPEC, and the other one is from the NGSL. As depicted in Figure \ref{Fig3}, the systematic errors for the high-extinction sources (represented by the red lines) in both pairs are very different from those for the low-extinction sources (represented by the blue lines), after excluding the effects of $BP-RP$ and $G$. This effect is most significant in the wavelength range of $336-400$\,nm. The problems at the short wavelengths are likely attributed to stray or scattered out-of-band light, a problem for all spectral dispersing elements (e.g., see point 9 of Section 5 in \citealt{Bohlin2019}).
The effect of extinction is not hard to understand, given that the errors depend on the SED. To accurately describe the SED, it is necessary to consider at least the intrinsic color and extinction, while disregarding the weaker effects of \logg \,and \feh.

\begin{figure*}[ht]
\centering
\includegraphics[width=160mm]{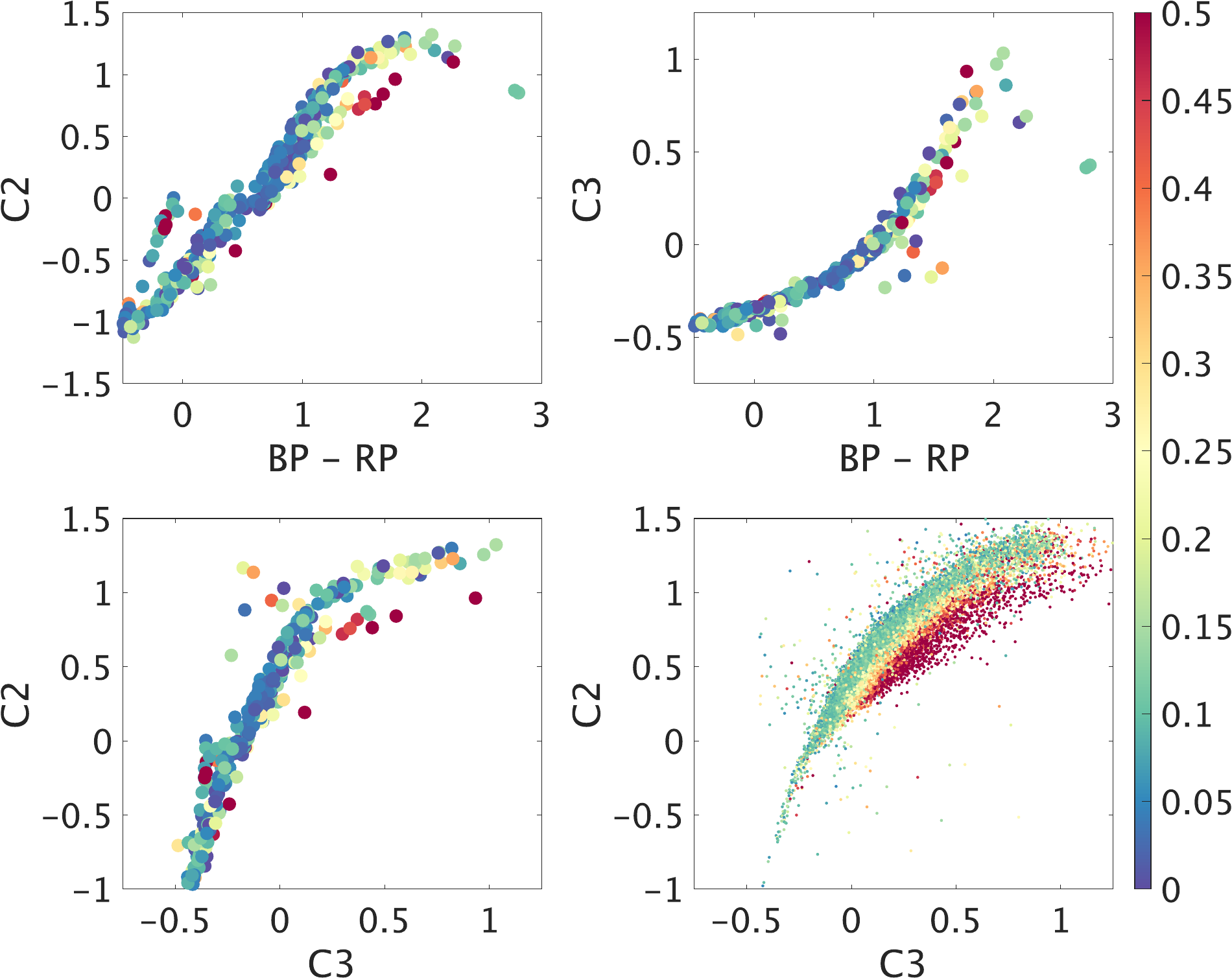}
\caption{Color-color diagrams of the training sample for $BP-RP$ vs. $C2$ (upper left), $BP-RP$ vs. $C3$ (upper right), and $C2$ vs. $C3$ (lower left). 
The lower right panel is identical to the lower left panel, but shows sources from the \texttt{XpSampledMeanSpectrum\_000000-003111.csv.gz} dataset ($G<15$). The color in all panels represents the \ebv~value.} 

\label{Fig4}
\end{figure*}

Given that $BP-RP$ and $E(B-V)$ represent the first two principal components of normalized stellar SEDs, with $E(B-V)$ being an indirect observable, we define two straightforward ``colors'' based on the slopes in the $390-476$\,nm and $478-600$\,nm wavelength ranges of the XP spectra as new principal components. Note here the XP spectra have been divided by the mean flux within the $460-490$\,nm range. The resulting ``colors'', denoted as $C2$ and $C3$ (the second and third segments of an XP spectrum) are scaled up by a factor of 100 to align numerically with $BP-RP$. Figure \ref{Fig4} displays that the combination of $C2$ and $C3$ can not only capture the color information of the sources but also facilitate the differentiation of their extinctions. The sources in the lower right panel are from \texttt{XpSampledMeanSpectrum\_000000-003111.csv.gz}\footnote{http://cdn.gea.esac.esa.int/Gaia/gdr3/Spectroscopy/ \\ xp\_sampled\_mean\_spectrum/} dataset and \ebv $\,$values are from SFD98.

In addition, a 0.2 arcsec slit was applied for nearly all the sources in NGSL. Any miscentering of the sources may cause errors in the relative flux distribution (\citealt{NGSL2009}). To mitigate the impact of this caveat on our corrections for the XP spectra, the NGSL sources in the training sample are scaled to the CALSPEC using the median in the range of $460-490$\,nm. The details of the scaling process are presented in the coming step.

\begin{equation} 
Systematic \,\, Error = f_1(C2) \times f_2(C3) \times f_3(G)
\label{SE2}
\end{equation}

For each wavelength, we utilize a local regression using weighted linear least squares along with a second-degree polynomial model to fit the systematic errors in orders of $f_1(C2)$, $f_3(G)$, and $f_2(C3)$, as shown in  Equation \ref{SE2}. 
In the regression process, our initial weights are proportional to the negative square of the relative errors of the Gaia XP spectra. Then, we increase the weights of the CALSPEC spectra by a factor of 5 to mitigate the impact of the caveat of the NGSL spectra.

We first apply the above procedure to a spectral segment ranging from $460-490$\,nm. The median residuals obtained from this segment are then utilized to scale each of the NGSL sources, so that each  NGSL source is consistent with CALSPEC in the $460-490$\,nm range. 
The scale values range from approximately 0.6 to 1.9, with most values between 0.98 and 1.02.
From here on, the training sample comprises the CALSPEC spectra and the scaled NGSL spectra. 

\begin{figure*}[ht]
\centering
\includegraphics[width=180mm]{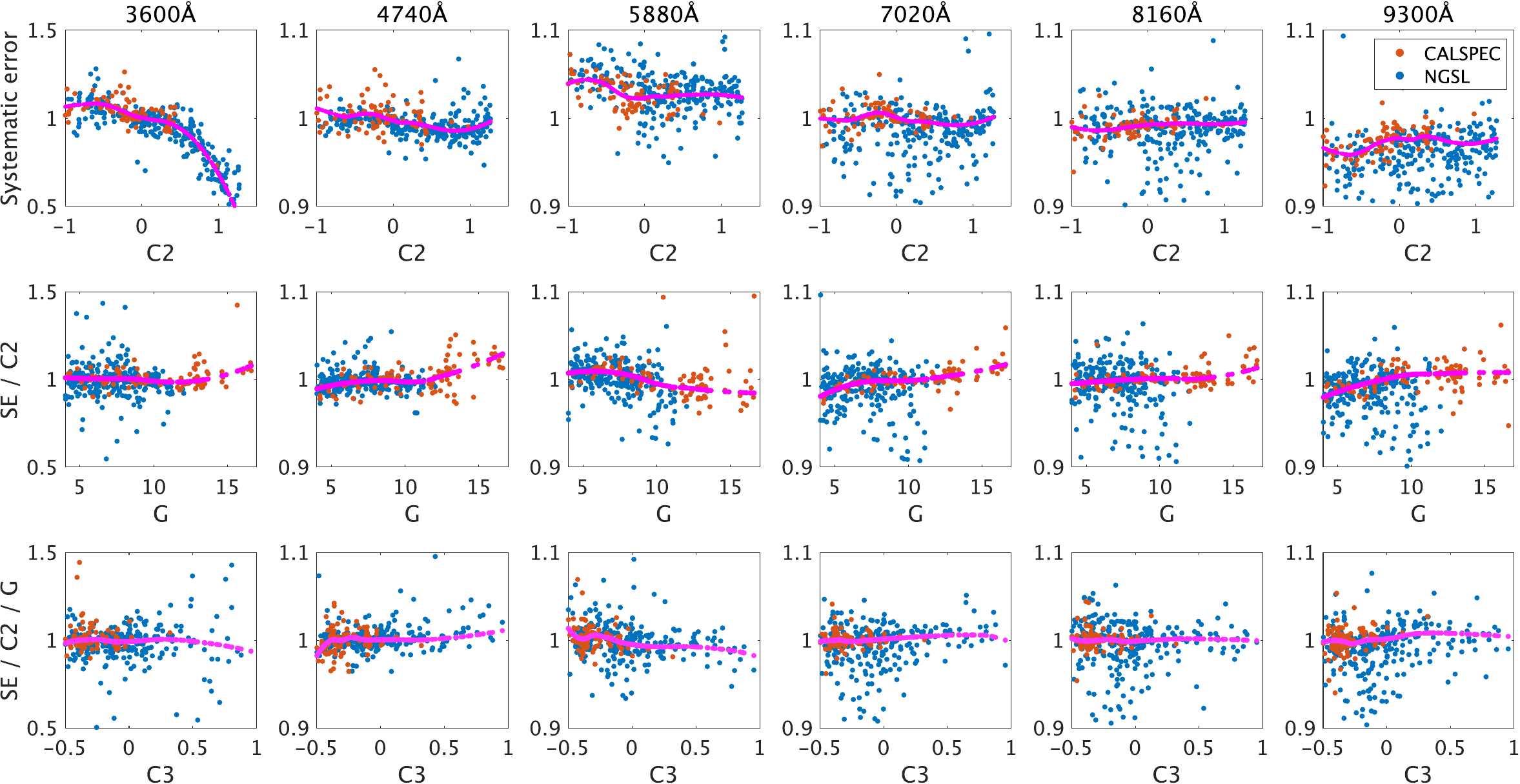}
\caption{The nontruncated XP spectra are used as an example to demonstrate the first iteration of the procedure. Red points represent sources from the CALSPEC, blue points correspond to sources from the NGSL, and magenta points depict the results of the fit at each step.} 
\label{Fig5}
\end{figure*}

\begin{figure*}[ht]
\centering
\includegraphics[width=180mm]{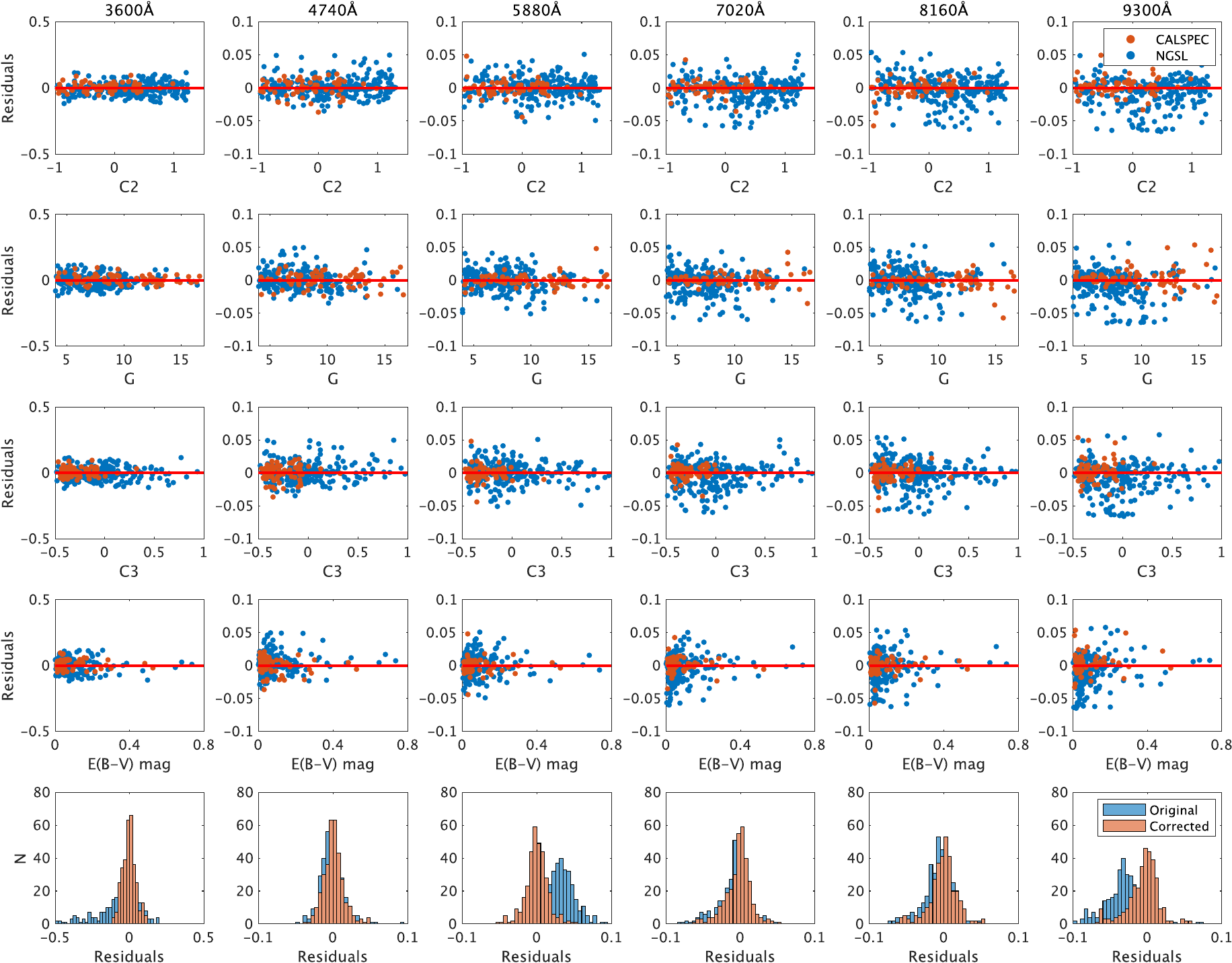}
\caption{Fitting residuals for the nontruncated XP spectra. The top four rows show the residuals as a function of $C2$, $G$, $C3$, and $E(B-V)$, respectively. The last row illustrates the comparison between the residuals and the original systematic errors, where the original systematic errors are subtracted by 1 to facilitate comparison. The red and blue points represent sources from the CALSPEC and NGSL, respectively.} 
\label{Fig6}
\end{figure*}

\begin{figure*}[ht]
\centering
\includegraphics[width=180mm]{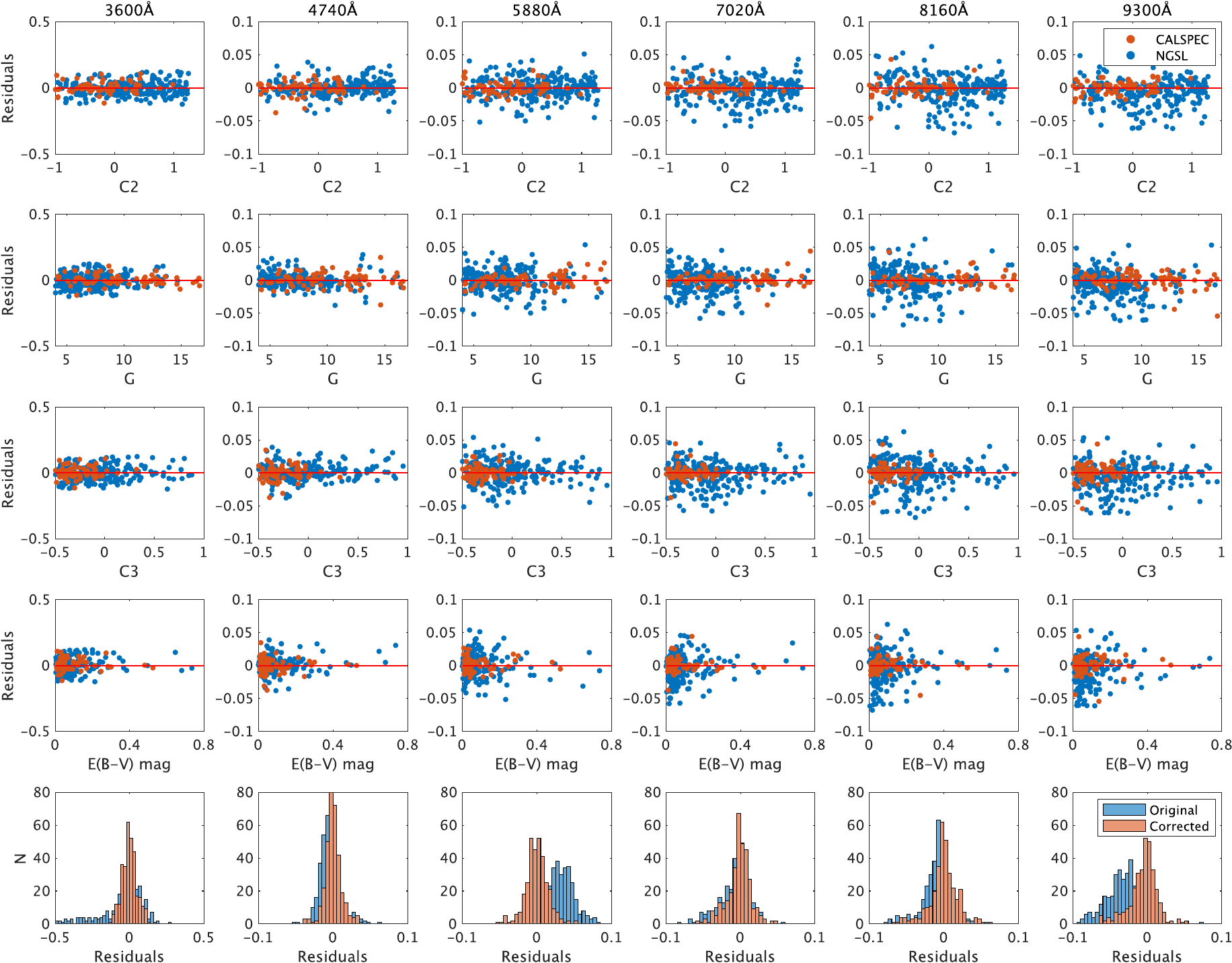}
\caption{Same as Figure \ref{Fig6} but for the truncated XP spectra.} 
\label{Fig7}
\end{figure*}

One iteration is performed and the procedure is depicted in Figure \ref{Fig5}. Then, we apply this procedure to all the wavelengths. The fitting residuals for some selected wavelengths are presented in Figure \ref{Fig6} and Figure \ref{Fig7} for the nontruncated and truncated XP spectra, respectively. 
The fitting residuals exhibit no trends with respect to $C2$, $G$, $C3$, and $E(B-V)$ within the parameter range of the training sample.

\begin{figure}[ht]
\includegraphics[width=80mm]{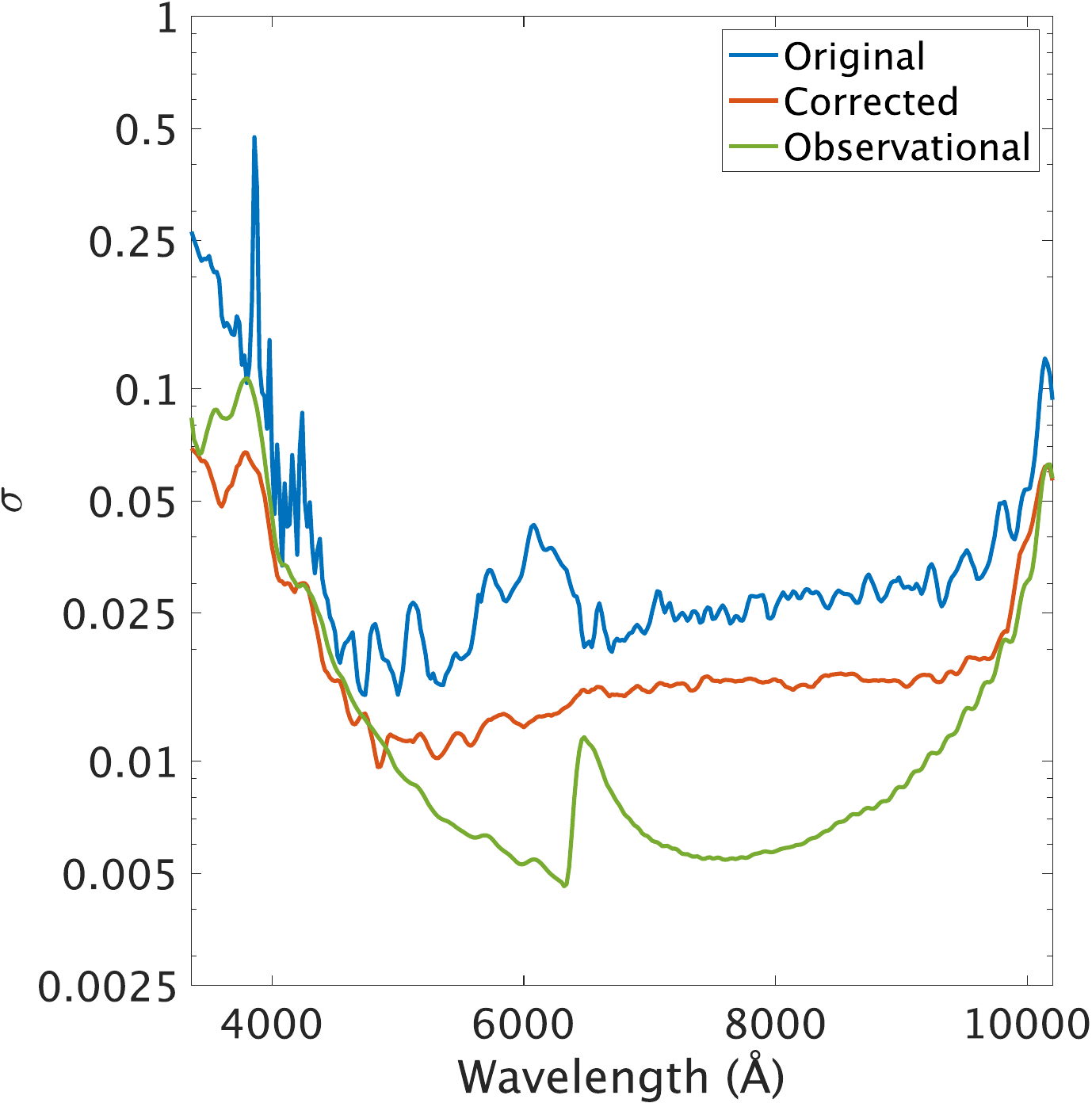}
\caption{The original systematic error (blue), dispersion of the fitting residuals (red), and typical observational error (green) of the training sample as a function of wavelength. } 
\label{Fig8}
\end{figure}

Figure \ref{Fig8} depicts the original systematic errors (blue line) and the dispersion of the fitting residuals (red line) with respect to wavelength. Given the similarity in results between the nontruncated and the truncated XP spectra, only the results for the former are shown.
The original systematic errors are estimated to be the standard deviation of the ratios between the original Gaia XP spectra and the training sample (excluding the largest and smallest 1.5 per cent outliers). 
When calculating the dispersion of the fitting residuals, a 2$\sigma$ clipping is performed. The median relative errors of the Gaia XP spectra of common sources are also depicted by the green line. Note that the relative errors depend on the $G$ magnitude and do not follow a normal distribution. 
The fitting residuals exhibit a dispersion of 2--7 per cent when the wavelength is shorter than $400$\,nm or longer than $960$\,nm, 
A comparison with the green line suggests that the aforementioned dispersion is primarily influenced by observational errors of the XP spectra.
 
At other wavelengths, the dispersion is around 1--2 per cent.
The discontinuity at approximately $635$\,nm in the blue line can be attributed to systematic errors in the conjunction between the BP and RP spectra. 
After our correction, this discontinuity disappears in the red line.

Until now, we have established the relation between the source properties and the systematic errors for the XP spectra. This implies that, for a given XP spectrum ($C2$, $C3$, and $G$) within the parameter space of the training sample, we are able to correct its systematic errors. And the three correction terms of $C2$, $C3$, and $G$ can be used independently or combined by multiplication. 

\section {An Updated $G$ Term Correction by LAMOST DR7}

As illustrated in Figure \ref{Fig1}, the correction of the $G$ term from the training sample may be unreliable when $G > 14$ due to the very limited number of faint sources. In this section, we utilize a revised version of the stellar color regression (SCR; \citealt{SCR2015a}, \citealt{2022S82}) method to provide an updated correction for the $G$ term within the range of $10<G<17.5$.

\subsection{Data selection}
There are about 5.7 million common sources between LAMOST DR7 and {\it Gaia} EDR3, using a matching radius of 1 arcsec. We consider the following criteria to ensure the data quality and facilitate the implementation of the SCR method:
\begin{enumerate}
    \item \texttt{duplicated\_source = False}
    \item the signal-to-noise ratios for the $g$ band of LAMOST spectra are greater than 30
    \item absolute values of Galactic latitude are greater than 20\degr 
    \item vertical distances to the Galactic disk are greater than 0.2 kpc
    \item $E(B-V)_{ \rm SFD } < 0.025$ 
    \item $5500 <\teff < 6500 \,K$
    \item $4 <$ \logg $\,< 4.5$
    \item $\feh > -1$
\end{enumerate} 

The SCR method in this work involves two sets of samples: a control sample to establish the relationship between the stellar atmospheric parameters and the normalized intrinsic Gaia XP spectra (which will be explained in detail later), and a target sample to update the $G$ term correction by applying this relationship.
In the case of the control sample, an additional criterion of $12<G<13$ is applied to mitigate the influence the $G$ term systematic errors of the XP spectra.
In addition, we adopt a second-order two-dimensional polynomial to fit the excess factor as a function of $BP-RP$ and \feh. We use a 3$\sigma$ clipping method (as described in \citealt{Xushuai}) to remove sources that are affected by background and contamination issues in the $BP$ and $RP$ bands for both the control and target samples.
After the above constraints, the control sample has approximately 40,000 sources and the target sample has approximately 220,000 sources.

\subsection{Method}
We assume that an XP spectrum is comprised of three components as described in Equation \ref{Gterm}: the intrinsic spectrum of the source, the interstellar extinction part, and the systematic errors. The systematic errors encompass the $f_1(C2)$ and $f_2(C3)$ terms and the $f_3(G)$ term. 
The intrinsic spectrum can be predicted using the SCR method based on the atmospheric parameters of the source. Note that, the SCR method cannot directly predict the magnitudes or fluxes. Therefore, we perform a normalization of the XP spectra at a later stage. 
The interstellar extinction can also be estimated (which will be discussed in detail later), and the $f_1(C2)$ and $f_2(C3)$ terms 
can be adopted from previous section.
We deduct the aforementioned components of the Gaia XP spectra that are either predictable or known. Consequently, the remaining $f_3(G)$ term can be estimated. 

\begin{gather}
Gaia \,\, XP \,\, spectrum = \notag \\
 Intrinsic \,\, spectrum \,\, (Atmospheric \,\, parameters) \notag \\
 \times Interstellar \,\, extinction \,\, (Interstellar  \,\,extinction \,\, model) \notag \\
 \times Systematic \,\, errors \,\, (f_1(C2) \times f_2(C3) \times f_3(G)) 
\label{Gterm}
\end{gather}

In order to remove the effect of extinction from XP spectra, we utilize the \ebv ~from the SFD98 map (\citealt{SFD98}) along with the Fitzpatrick \rv$\,\,= 3.1$  extinction curve (F99; \citealt{F99}). 
The 14 per cent overestimation of reddening  in the SFD98 map (\citealt{Schlafly2011};\citealt{Yuan2013}) has also been taken into account.
Then we divide the dereddened XP spectra by the mean flux within the $460-490$\,nm range, so that they can be predicted by atmospheric parameters (\teff, \logg, and \feh). 

\begin{figure*}[ht]
\centering
\includegraphics[width=180mm]{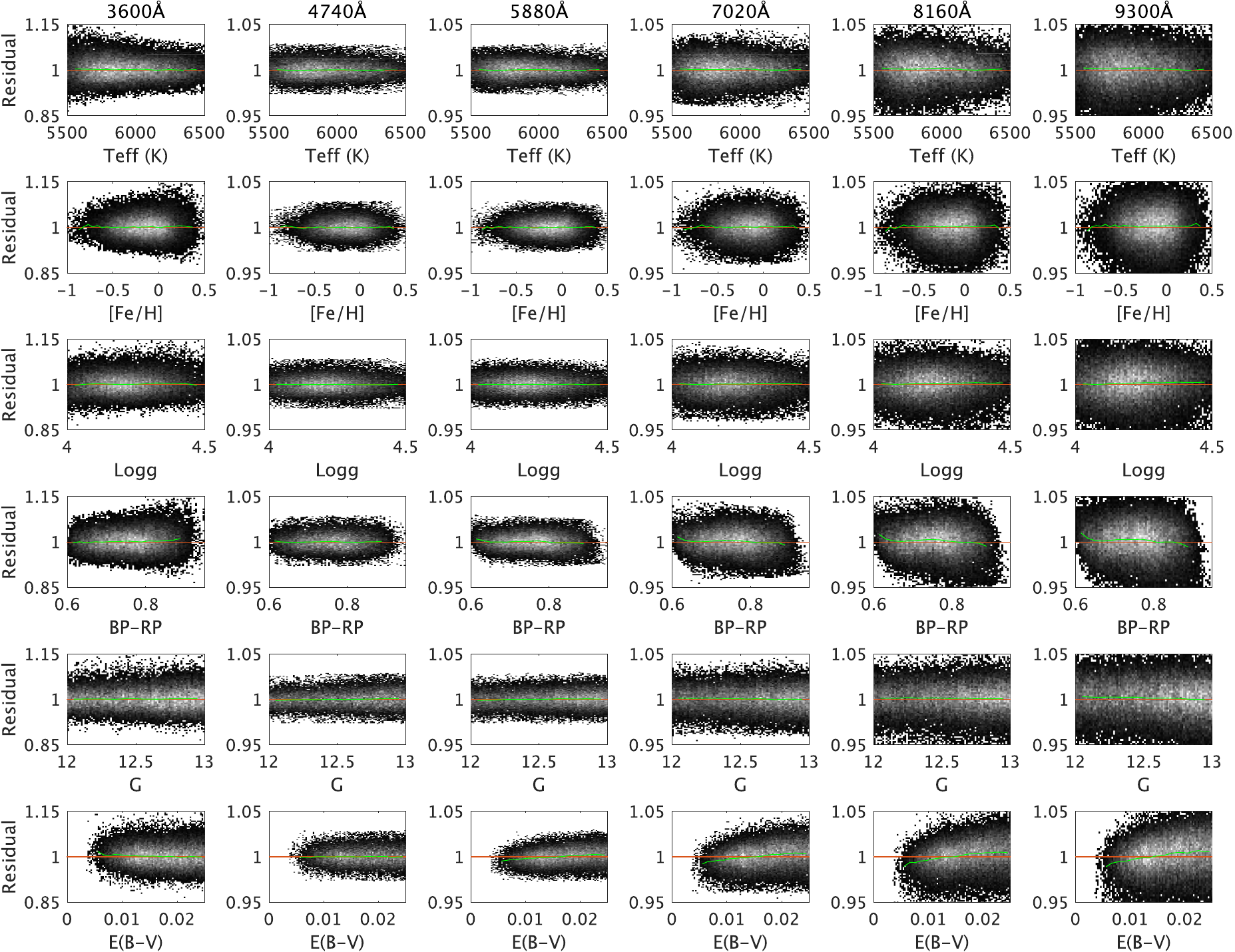}
\caption{Fitting residuals of the SCR method for the control sample. From top to bottom: \teff, \feh, \logg, $BP-RP$, $G$ and \ebv $\,$ respectively. The green line indicates the median value of residuals.} 
\label{Fig9}
\end{figure*}

For the control sample, the dereddened and normalized XP spectra are fully corrected by the $C2$, $C3$ and $G$ terms and subsequently fitted as a function of ${\teff}$, \logg, and ${\feh}$ using the following function:

\begin{gather}
    F = a_1 \times \teff ^3 + a_2 \times \teff ^2 \times \feh + a_3 \times \teff \times \feh ^2 \notag \\
    + a_4 \times \feh ^3 + a_5 \times \teff ^2 + a_6 \times \teff \times \feh \notag \\
    + a_7 \times  \feh ^2 + a_8 \times \teff + a_9 \times \feh \notag \\
    + a_{10} \times \logginequ ^3 + a_{11} \times \logginequ ^2 + a_{12} \times \logginequ  + a_{13}
\label{F1}
\end{gather}
For each wavelength,  a set of coefficients from $a_1$ to $a_{13}$ is obtained through the fitting process. The residuals for specific wavelengths are depicted in Figure \ref{Fig9} for the nontruncated XP spectra. The truncated XP spectra exhibit similar results. For red wavelengths ($\lambda > 600$\,nm), the residuals exhibit a weak correlation with \ebv, likely caused by the systematic errors in the F99 extinction curve.

\begin{figure*}[ht]
\centering
\includegraphics[width=180mm]{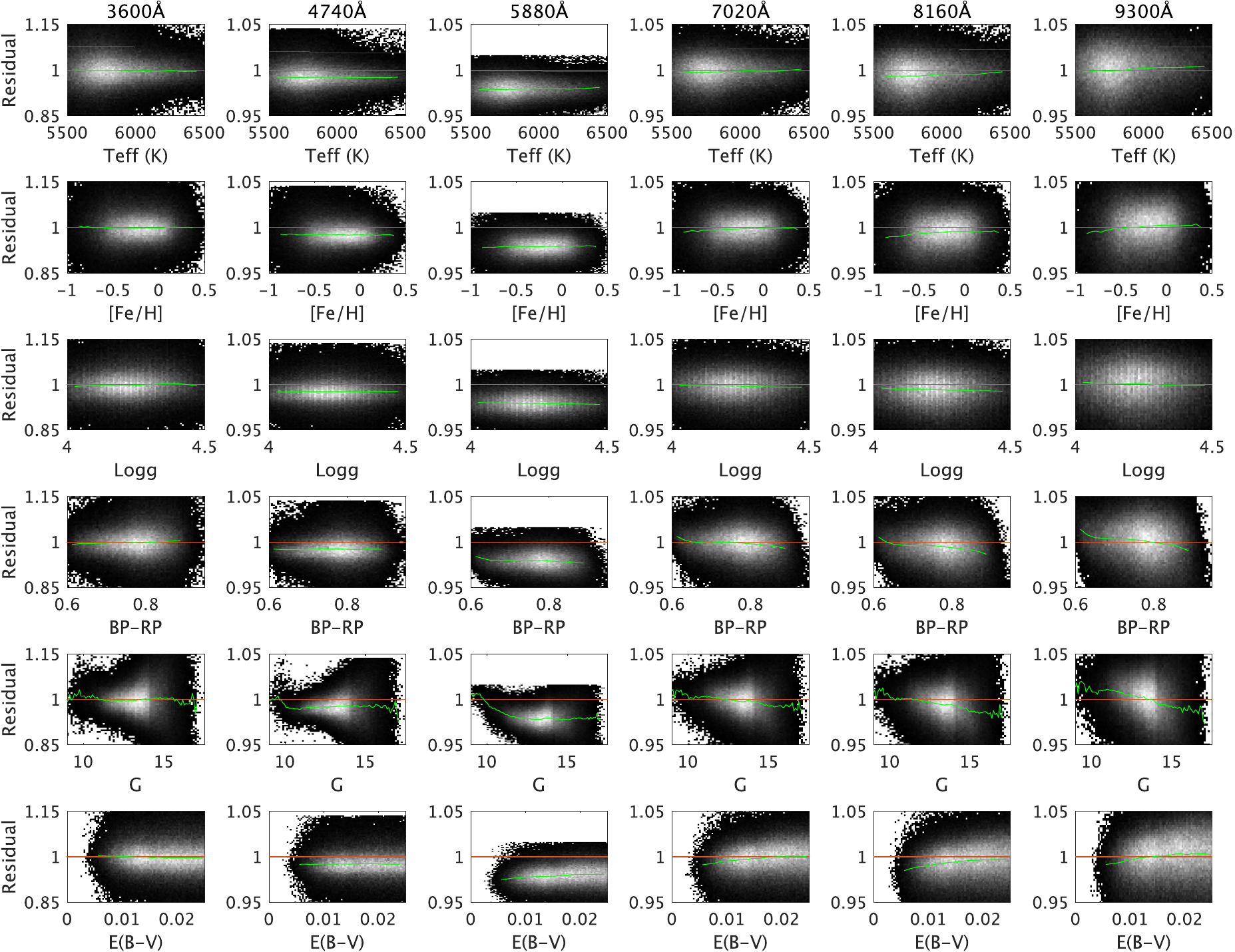}
\caption{Similar to Figure \ref{Fig9}, 
the remaining differences to illustrate the systematic errors in the $f_3(G)$ term. See more details in Section\,4.} 
\label{Fig10}
\end{figure*}

For the target sample, the dereddened and normalized XP spectra are firstly corrected by the $f_1(C2)$, $f_2(C3)$ terms on one hand. On the other hand, Equation \ref{F1} is used to deduce their intrinsic, normalized spectra. Then, any remaining differences can be attributed to systematic errors in the $f_3(G)$ term. The differences for some selected wavelengths are illustrated in Figure \ref{Fig10} for the nontruncated spectra. The truncated XP spectra exhibit similar results. 

We employ the $f_3(G)$ corrections given by the smoothed green lines in the fifth row of Figure \ref{Fig10} for the magnitude range $10<G<17.5$. Due to the corrections based on the SCR method being computed for the normalized XP spectra, 
we further multiply this result by the mean corrections within the wavelength range of $460$\,nm to $490$\,nm, as determined from the training sample,  to do the $f_3(G)$ corrections of the Gaia XP spectra.
For the $G$ magnitude range beyond the training sample ($G>16.5$), the mean corrections at $G=16.5$ are adopted.

\section {Analyses and tests of the correction}
\subsection{Analysis}
\begin{figure*}[ht]
\centering
\includegraphics[width=180mm]{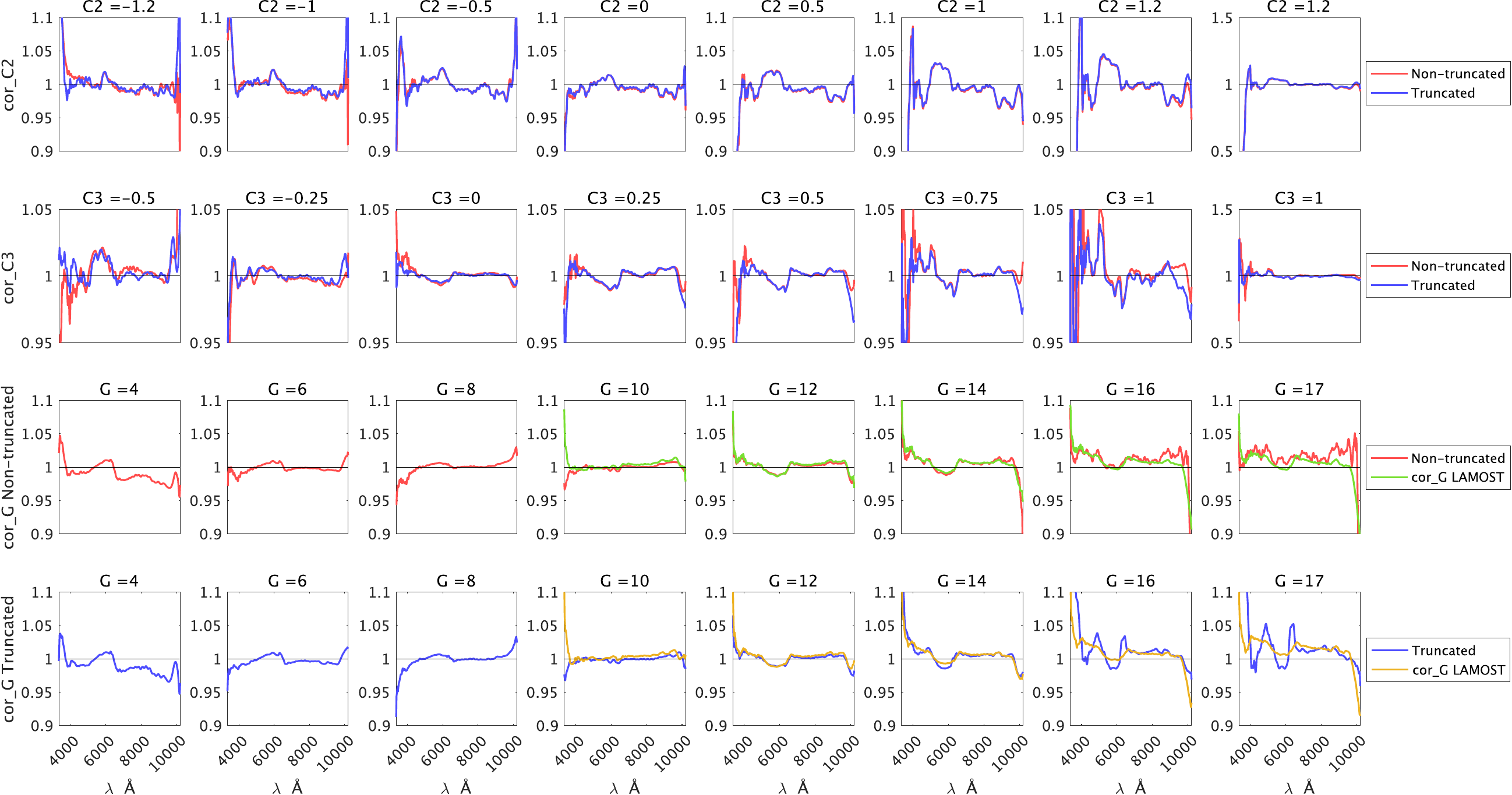}
\caption{The correction of the nontruncated (red lines) and truncated (blue lines) XP spectra in $C2$, $C3$ and $G$ term.  The green and yellow lines represent the additional correction in the $G$ term for the nontruncated and truncated XP spectra, respectively.}
\label{Fig11}
\end{figure*}

The results obtained from the two preceding sections have been integrated to provide a comprehensive correction for the XP spectra, which is depicted in Figure \ref{Fig11}. The corrections are more significant toward the blue end ($336-400$\,nm), and
are less than 5 per cent elsewhere.  Additionally, all correction terms exhibit a jump at $630 $\,nm, which is attributed to the junction of the BP and RP spectra at this point. 
The corrections for the nontruncated and truncated XP spectra are generally consistent, with the exception of some variations in the $C3$ term. Concerning the additional correction for the $G$ term based on LAMOST, it does not consist with the correction provided by the training sample at $G = 16$. This inconsistency is due to the limited number of sources fainter than $G = 14$ in the training sample. The observed discrepancy in the blue end at $G=10$ may be attributed to the lack of blue stars of the training sample within this range, as shown in Figure \ref{Fig1}. When $G > 10$, the additional correction in the $G$ term is preferred to the correction derived from the training sample in the final correction.

\begin{figure*}[ht]
\centering
\includegraphics[width=180mm]{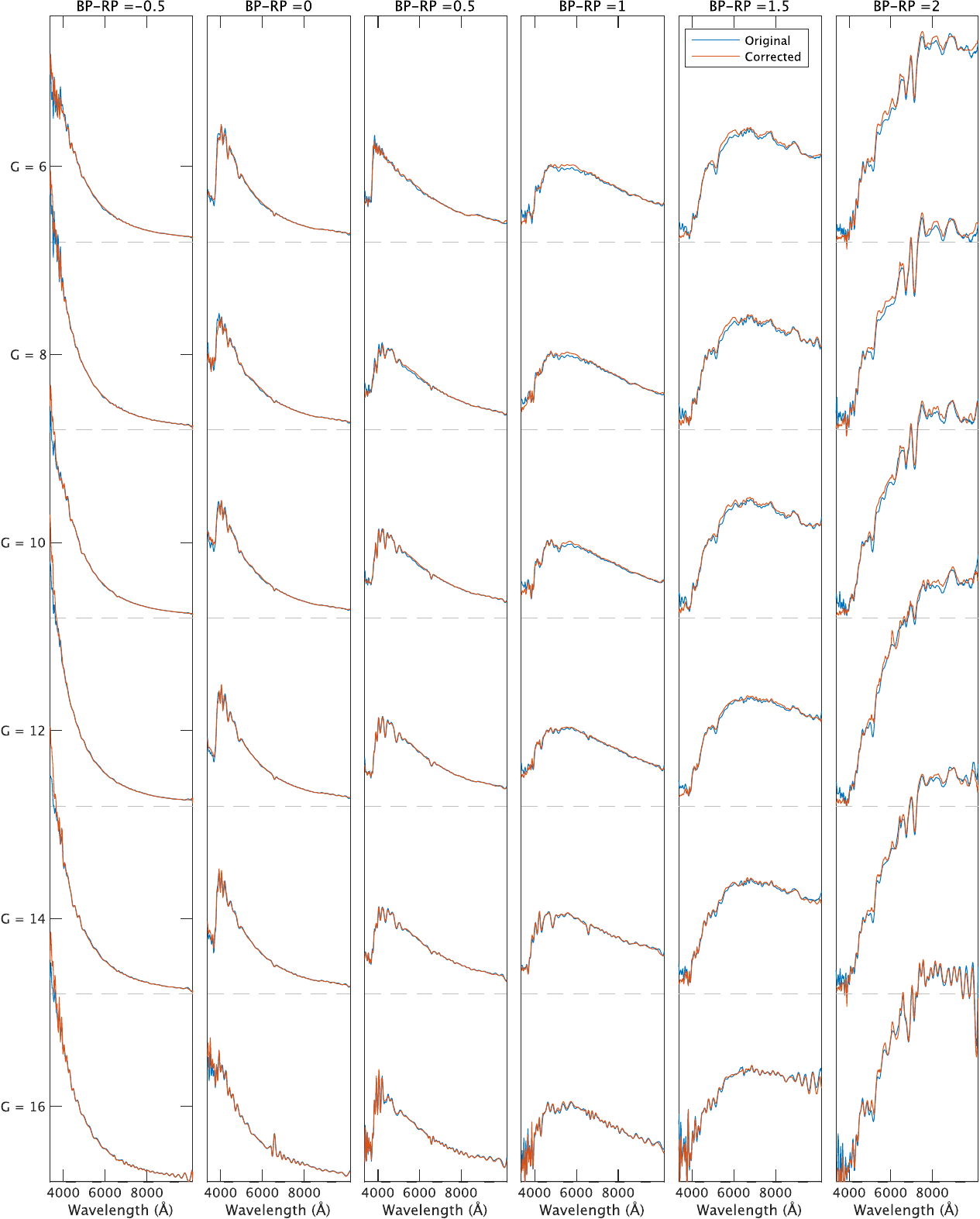}
\caption{The comparison between the original (blue lines) and corrected (red lines) nontruncated XP spectra for sources of different colors and magnitudes. The grey horizontal dashed lines indicate the positions where the flux of each spectrum is zero.} 
\label{Fig12}
\end{figure*}

We employ the nontruncated XP spectra in Figure \ref{Fig12} to illustrate our correction outcomes for sources of various colors and magnitudes. 
As depicted in Figure \ref{Fig12}, our corrections exhibit significant improvements, particularly for wavelengths below $400$\,nm.

\subsection{Test}
\begin{figure}[ht]
\centering
\includegraphics[width=80mm]{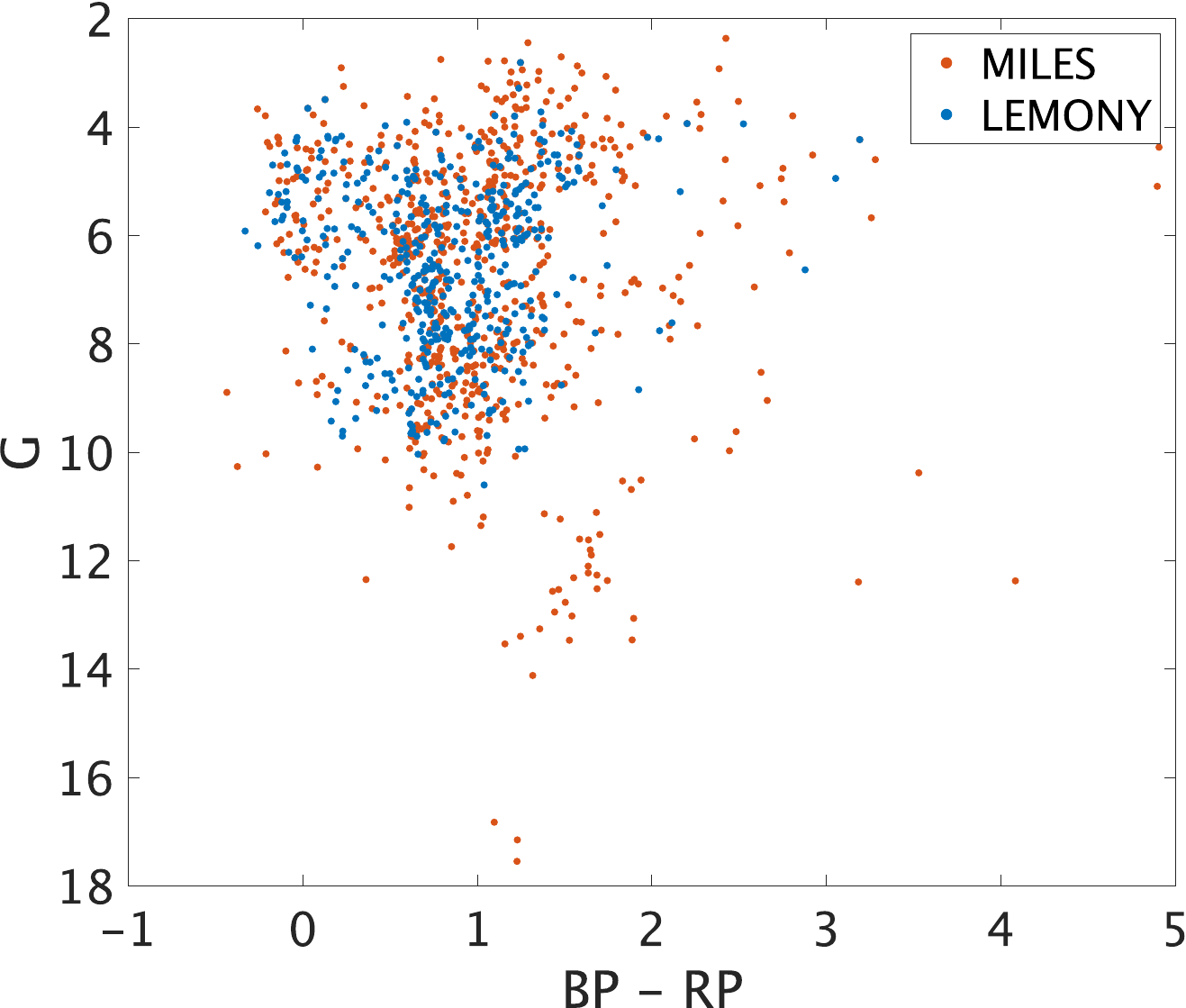}
\caption{The color-magnitude diagram of the test sample. Red points represent sources from MILES while blue points correspond to sources from LEMONY.} 
\label{Fig13}
\end{figure}

The MILES (\citealt{MILES}) and LEMONY (\citealt{LEMONY}) spectra (hereafter the test sample) are utilized to verify our corrections. A total of 725 common sources are obtained by cross-matching the MILES and Gaia XP spectra, with a maximum angular separation of 5 arcsec.
Regarding the LEMONY, 453 common sources with Gaia XP spectra are identified using the same cross-matching method as MILES.
Properties of the test sources are presented in Figure \ref{Fig13}.

Similar to the CALSPEC and NGSL spectra, we convolve the MILES and LEMONY spectra to match the resolution of the XP spectra. We then interpolate the test sample to the sampled wavelength of the XP spectra using a cubic spline. It is noteworthy that the MILES and LEMONY spectra were dereddened using the F99 \rv$\,\,= 3.1$ extinction curve and \ebv~ provided by themselves. Therefore, the same dereddening approach is applied to the XP spectra to enable comparison with the test sample.
The MILES spectra are normalized by the mean flux between the wavelengths of $442-458$\,nm, along with the corresponding XP spectra. Some additional processes are necessary for the LEMONY spectra due to issues with the conjunction of spectra from two spectrographs, as well as uncorrected telluric absorption features in the red region of the spectra. 
Firstly, we normalize the blue region of the spectra using the mean flux between the wavelengths of $442-458$\,nm, following the same method as MILES. Subsequently, the red region is normalized using the mean flux between $588-608$\,nm. Prior to the convolution, strong telluric absorption features ($685-704$\,nm, $715-740$\,nm, $758-773$\,nm and $805-837$\,nm) are clipped by linear interpolation. In the subsequent comparison, fluxes at the conjunction region between the two spectrographs ($510-520$\,nm) are masked.

\begin{figure*}[ht]
\centering
\includegraphics[width=160mm]{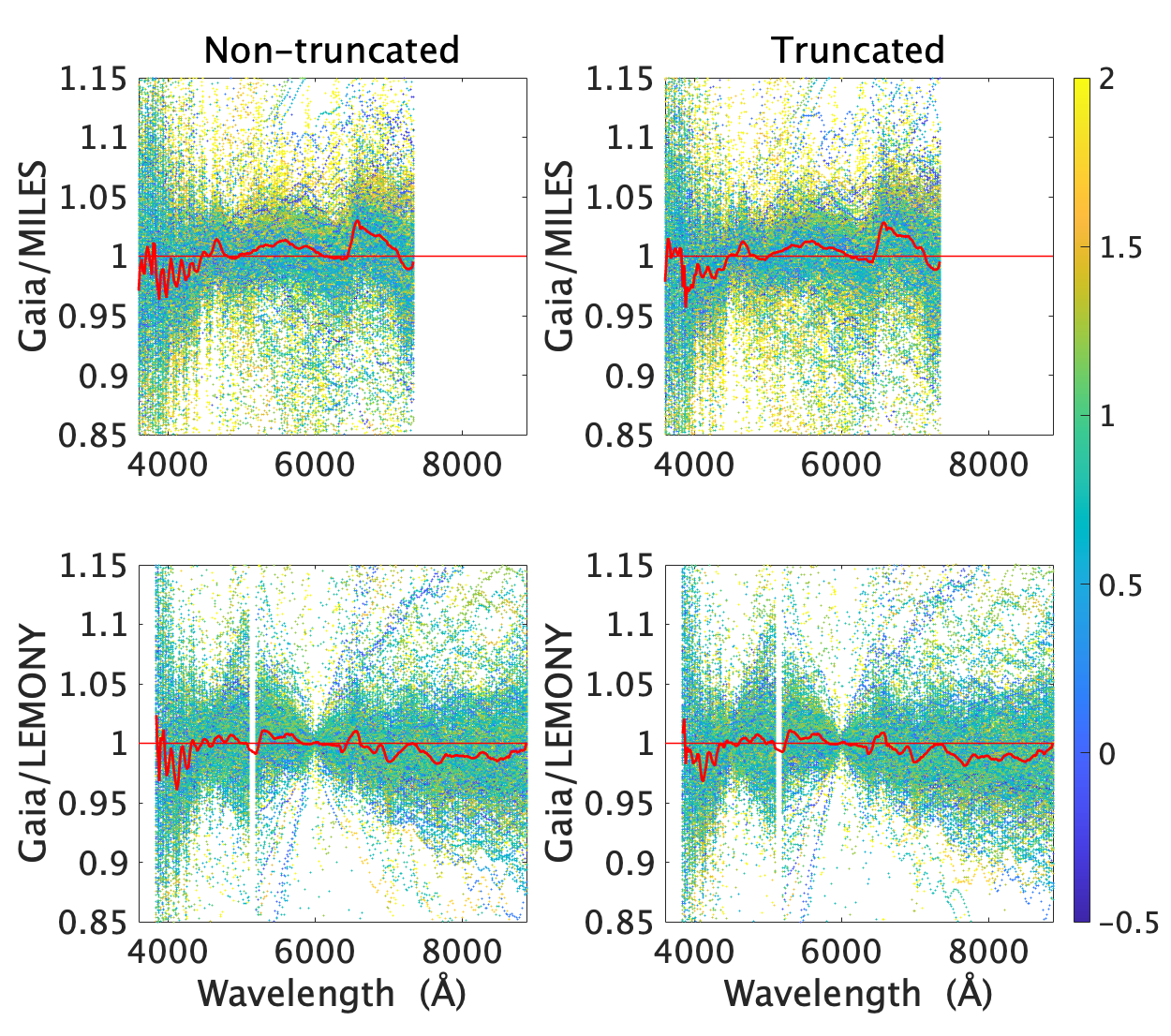}
\caption{A direct comparison between the corrected XP spectra and the test sample. The color of each point represents the $BP-RP$ color of the corresponding source, and the solid red line denotes the median value of the differences at each wavelength.} 
\label{Fig14}
\end{figure*}

\begin{figure}[ht]
\includegraphics[width=80mm]{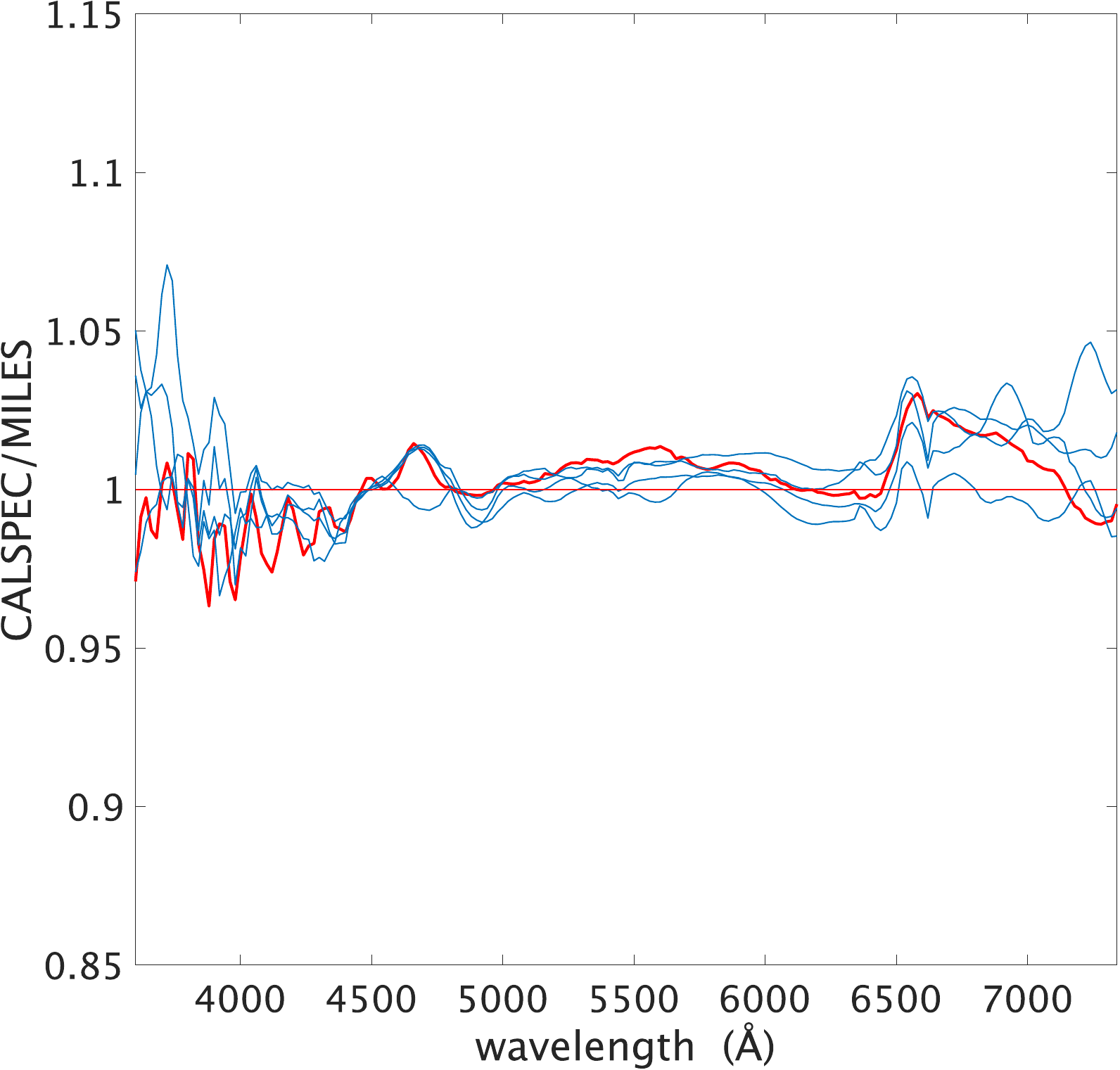}
\caption{Comparison results between the nontruncated corrected XP spectra and MILES (shown as the red line), as well as the comparison between the four common sources of CALSPEC and MILES (depicted as the blue lines).} 
\label{Fig15}
\end{figure}

Figure \ref{Fig14} displays the ratios of the test sample and their corresponding XP spectra, with two notable features worth mentioning. Firstly, the ratios of MILES and XP spectra exhibit a jump at approximately $650$\,nm, a known issue of MILES which is also observed in the comparison between the MILES and the LEMONY spectra (refer to Figure 12 in \citealt{LEMONY}). It is attributed to imperfect correction of second-order contamination (\citealt{MILES}) in the MILES spectra. Secondly, there are discrepancies up to about 3 per cent in the blue around 400\,nm for both the MILES and LEMONY spectra. Such discrepancies are partly attributed to the differences between the training and test samples.
There are four common sources between the CALSPEC and the MILES, and their comparison in Figure \ref{Fig15} shows a similar pattern to Figure \ref{Fig14}.

\section{Absolute calibration to the corrected Gaia DR3 photometry}

Based on the result of the previous sections, it is evident that our corrected Gaia XP spectra exhibit consistency with the training sample. 
In this section, we perform an absolute calibration of our result by comparing the synthetic photometry of our corrected XP spectra to the corrected Gaia DR3 photometry (\citealt{YangLinEDR3}), so that there are no magnitude-dependent systematic errors in the range of $10<G<17.65$. 

\cite{YangLinEDR3} have carried out an independent validation and correction of Gaia DR3 photometry by using approximately 10,000 Landolt standards (\citealt{landolt2013}). Trends of up to 10 mmag respect to $G$ magnitude have been identified for the $G$, $BP$, and $RP$ magnitudes within the range of $10 < G < 19$. Their absolute zero-point corrections are derived from the synthetic magnitudes computed using the CALSPEC spectra (see their Figure\,4).  
We revise the $BP$ absolute zero-point of \cite{YangLinEDR3}, and further details can be found in Appendix.

The Landolt standards rely on ground-based observations, where the CCD detectors are not subject to damage from cosmic rays and trapped particle radiation in the Van Allen belts, unlike the STIS detector (\citealt{CALSPEC22}).
Consequently, the complex correction for the charge transfer efficiency (CTE) implemented in the CALSPEC (\citealt{CALSPEC22}) is unnecessary for the Landolt standards. And the CTE correction applied in CALSPEC depends on brightness.
In addition, the nonlinearity effect of CCD detectors is relatively easy to compensate in Landolt's observation (see their Appendix A and B; \citealt{landolt2013}). Furthermore, the variable atmospheric contribution to the passband function does not result in magnitude-dependent systematic errors.
Therefore, we favor the belief that Landolt standards are more likely to be accurate if there are magnitude-dependent differences between the CALSPEC and Landolt stars.
Furthermore, given the insufficiency of faint sources in the CALSPEC, the corrected Gaia DR3 photometry proves to offer better absolute zero-points in this brightness range.

\subsection{Result and analysis}
The absolute calibration is conducted using sources from \texttt{XpSampledMeanSpectrum\_000000-003111.csv.gz}, the same as those in the lower right panel of Figure\,\ref{Fig4}. To ensure the accuracy of $BP$ and $RP$ photometry, we also employ the criterion \texttt{phot\_bp\_rp\_excess\_factor} $<1.26+0.04 \times (BP-RP)^2$). 

\begin{figure*}[ht]
\includegraphics[width=180mm]{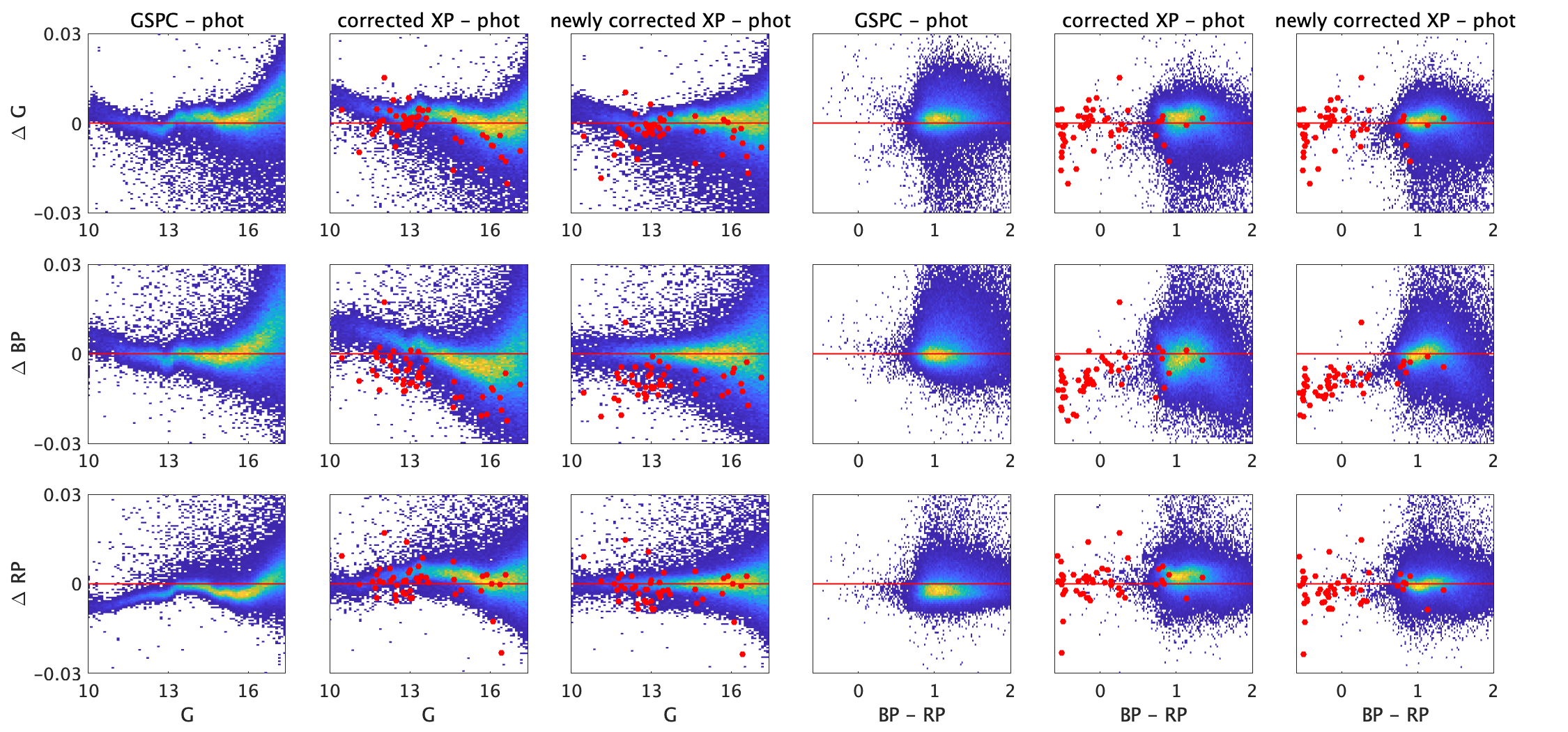}
\caption{Comparison of the synthetic photometry based on the XP spectra with the corrected Gaia DR3 photometry. 
The left and right three columns plot the differences as a function of $G$ and $BP-RP$, respectively.
The first/fourth, second/fifth,  and third/sixth columns depict, respectively the results of synthetic photometry based on the original XP spectra, the corrected XP spectra, and the corrected XP spectra with absolute calibration.
The color of the background indicates the number of sources. The red dots represent the sources from the CALSPEC. The horizontal red lines indicate zero differences.} 
\label{Fig16}
\end{figure*}

Firstly, we concentrate on the trends with respect to the $G$ magnitude.
The synthetic photometry based on the original Gaia XP spectra disagrees with the corrected Gaia DR3 photometry. The discrepancies depend on the $G$ magnitude, as seen in the first column of Figure\,\ref{Fig16}.
The synthetic photometry based on our corrected Gaia XP spectra partially reduces the discrepancies, especially for the BP band at $G \approx 13$, as illustrated in the second column of Figure\,\ref{Fig16}. However, there are still discrepancies left, which are likely attributed to the discrepancies between the CALSPEC spectra and the corrected Gaia DR3 photometry.

We subsequently compute the synthetic magnitudes of the CALSPEC sources using the CALSPEC spectra and compare them with their corresponding corrected Gaia DR3 photometry, depicted as red dots. Very similar trends between the red dots and the contours with the $G$ magnitude are observed. 
Note the trends are different in the $G$, $BP$ and $RP$ bands, which are potentially attributed to the different systematic errors with respect to $G$ magnitude from G430L ($290-570$\,nm) and G750L ($524-1027$\,nm) regions of STIS. 
In addition, an upturn is observed at $G \approx 17$ in the $G$, $BP$ and $RP$ bands. This is attributed to the limited number of sources available in the CALSPEC spectra within this $G$ magnitude range, preventing us from effectively rectifying the systematic errors in the $G$ term.
More details are described at the end of Section\,4. 

To solve this issue, we perform an absolute calibration for our corrected Gaia XP spectra with the corrected Gaia DR3 photometry.
The absolute calibration consists of wavelength-independent adjustments within the wavelength ranges of $336-544$\,nm and $546-1020$\,nm in our correction. The adjustments are determined by the medians of the contours in the second and third row of the second column in Figure\,\ref{Fig16}.
Since the passband ranges for the $BP$ and $RP$ do not exactly match the aforementioned wavelength ranges, we have made the specific adjustments using the following equations:

\begin{gather}
    Adjustment_{G750L} = \Delta_{RP} \\
    Adjustment_{G430L} = \frac{\Delta_{BP}-y \times Adjustment_{G750L}}{x} \\
    x = \frac{\int _{325}^{544} T_{BP}(\lambda)\lambda d\lambda}{\int _{325}^{544} T_{BP}(\lambda)\lambda d\lambda + \int _{546}^{750} T_{BP}(\lambda)\lambda d\lambda} \\
    y = \frac{\int _{546}^{750} T_{BP}(\lambda)\lambda d\lambda}{\int _{325}^{544} T_{BP}(\lambda)\lambda d\lambda + \int _{546}^{750} T_{BP}(\lambda)\lambda d\lambda}  \notag \\
\end{gather}

In the above equations, $T_{BP}$ and $T_{RP}$ represent the transmission curves of the $BP$ and $RP$ bands, respectively. $\Delta_{BP}$ and $\Delta_{RP}$ denote the differences between the synthetic photometry of the corrected XP spectra and the corrected Gaia DR3 photometry in the $BP$ and $RP$ bands, respectively. 
$Adjustment_{G430L}$ and $Adjustment_{G750L}$ correspond to the adjustments for the wavelength ranges of $336-544$\,nm and $546-1020$\,nm, respectively.
Our final result is computed for the original Gaia XP spectra by employing the following equations, in which the systematic error indicates the correction from Section 3 to 4.

\begin{gather}
    Corrected \,\, XP\,_{336-544\,nm} \,\, =  \notag \\
    Orignal \,\, XP\,_{336-544\,nm} \notag \\
    \times Systematic \,\, Error  \notag \\
    \times 10^{\frac{Adjustment_{G430L}}{2.5}} \\
    Corrected \,\, XP\,_{546-1020\,nm} \,\,= \notag \\
    Orignal \,\, XP\,_{546-1020\,nm} \notag \\
    \times Systematic \,\, Error  \notag \\
    \times 10^{\frac{Adjustment_{G750L}}{2.5}} \notag \\
\end{gather}

The implementation of the absolute calibration to the corrected Gaia XP spectra leads to a better agreement between the synthetic photometry and the corrected Gaia DR3 photometry, the same as the CALSPEC sources. However, there is a slight inconsistency in the comparison of the $G$-band results when $G < 12$. This inconsistency is likely attributed to the correction to Gaia photometry from \cite{YangLinEDR3}, as there are few sources in this magnitude range (see their Figure\,3).

We plot the differences as a function of the $BP-RP$ color in
the three right columns of Figure\,\ref{Fig16}. 
For the synthetic photometry based on the original XP spectra (the fourth column), no systematic trends are found. 
This consistency is due to the fact that the processes applied to both the Gaia XP spectra and photometry are similar, resulting in similar systematic errors with respect to $BP-RP$. 
For the synthetic photometry based on the corrected XP spectra (the fifth and sixth columns), the trends, most obvious in the $BP$ band, are the results of our corrections, which have addressed the color-related systematic errors in the Gaia XP spectra.
Note that such color-dependent trends can explain the overall shift observed between the red dots and the contours in the second and third columns, i.e., the CALSPEC sources have much bluer colors compared to most sources used in the contours.

In conclusion, we incorporate this absolute calibration into our final corrections to ensure the absence of magnitude-dependent systematic errors within the range of $10<G<17.65$.

\section{Summary}
We have conducted a comprehensive correction of the Gaia DR3 XP spectra by combining data from CALSPEC and NGSL, as well as spectroscopic data from the LAMOST DR7.
We firstly compare the differences between the Gaia XP spectra and the training sample consisting of CALSPEC and NGSL spectra, under the assumption that the observed differences can be attributed to the systematic errors inherent in the Gaia XP spectra. 
We find that the systematic errors depend on not only
$BP-RP$ and $G$, but also on reddening \ebv.
We subsequently model the systematic errors as a function of colors $C2$, $ C3$ and the $G$ magnitude. 
The fitting residuals exhibit a dispersion of 1--2 per cent within the wavelength range of $400-900$\, nm and 2--7 per cent when the wavelength is shorter than $400$\,nm or longer than $960$\,nm, where the observational errors of the XP spectra are significantly increased and contribute most of the dispersion. 
This suggests that we have corrected the systematic errors in the Gaia XP spectra well, with an internal precision of 1--2 per cent.

Comparisons between the corrected Gaia XP spectra and the MILES and LEMONY spectra show that the $BP-RP$ color and $G$-magnitude-dependent systematic errors are no longer noticeable. 
The consistency between the corrected Gaia XP spectra and the MILES/LEMONY data is about 2 per cent in the $336-400$\,nm range and 1 per cent in other wavelengths, with a small systematic offset independent of $BP-RP$ and $G$ magnitude. 
The small offset mainly comes from the differences between the CALSPEC spectra and the MILES/LEMONY spectra.
The independent tests have demonstrated that our correction effectively reduces most systematic errors, especially in the near ultraviolet. 

We further perform an absolute calibration by comparing the synthetic photometry of our corrected XP spectra to the corrected Gaia DR3 photometry of \cite{YangLinEDR3}. The synthetic photometry from the final corrected Gaia XP spectra exhibits a good consistency with the corrected Gaia photometry.

A Python package is publicly available\footnote{https://doi.org/10.12149/101375 or https://github.com/HiromonGON/GaiaXPcorrection} to correct both the nontruncated and truncated sampled Gaia XP spectra within the range of approximately $-0.5<BP-RP<2$, $E(B-V)<0.8$, and $3<G<17.5$. This package will be continuously updated as new spectral libraries become available in the future\footnote{The current correction corresponds to version V1.0, trained using the 2023 September update of the CALSPEC dataset. 
A previous version (V0.1) is also available, 
which was trained using the 2020 April update of the CALSPEC dataset. 
Compared to the April 2020 update version, the
main update of the latest version of CALSPEC is the implementation of a new CCD nonlinear CTE correction (\citealt{CALSPEC22}).
The overall changes between versions V1.0 and V0.1 are not significant. Alterations to the $C2$ and $G$ terms are virtually unchanged. Modest changes have been observed only in the $C3$ term and within the $336-400$\,nm range.
The corrected Gaia XP spectra with version V0.1 have been employed in the photometric recalibration of the Javalambre Photometric Local Universe Survey (J-PLUS; \citealt{JPLUS}) and the Southern Photometric Local Universe Survey (S-PLUS; \citealt{SPLUS}), and the corrections have been confirmed (\citealt{XiaoJPLUS, XiaoSPLUS}).
}.

Our study not only opens up new possibilities for using XP spectra but also lays the groundwork for future investigations in many fields.

\appendix
\setcounter{secnumdepth}{0}
\setcounter{table}{0}   
\setcounter{figure}{0}
\renewcommand{\thetable}{A\arabic{table}}
\renewcommand{\thefigure}{A\arabic{figure}}

\section{A. Correction of the Absolute Zero-point in the $BP$ Band of Yang et al. (2021)}

\begin{figure*}[ht]
\includegraphics[width=160mm]{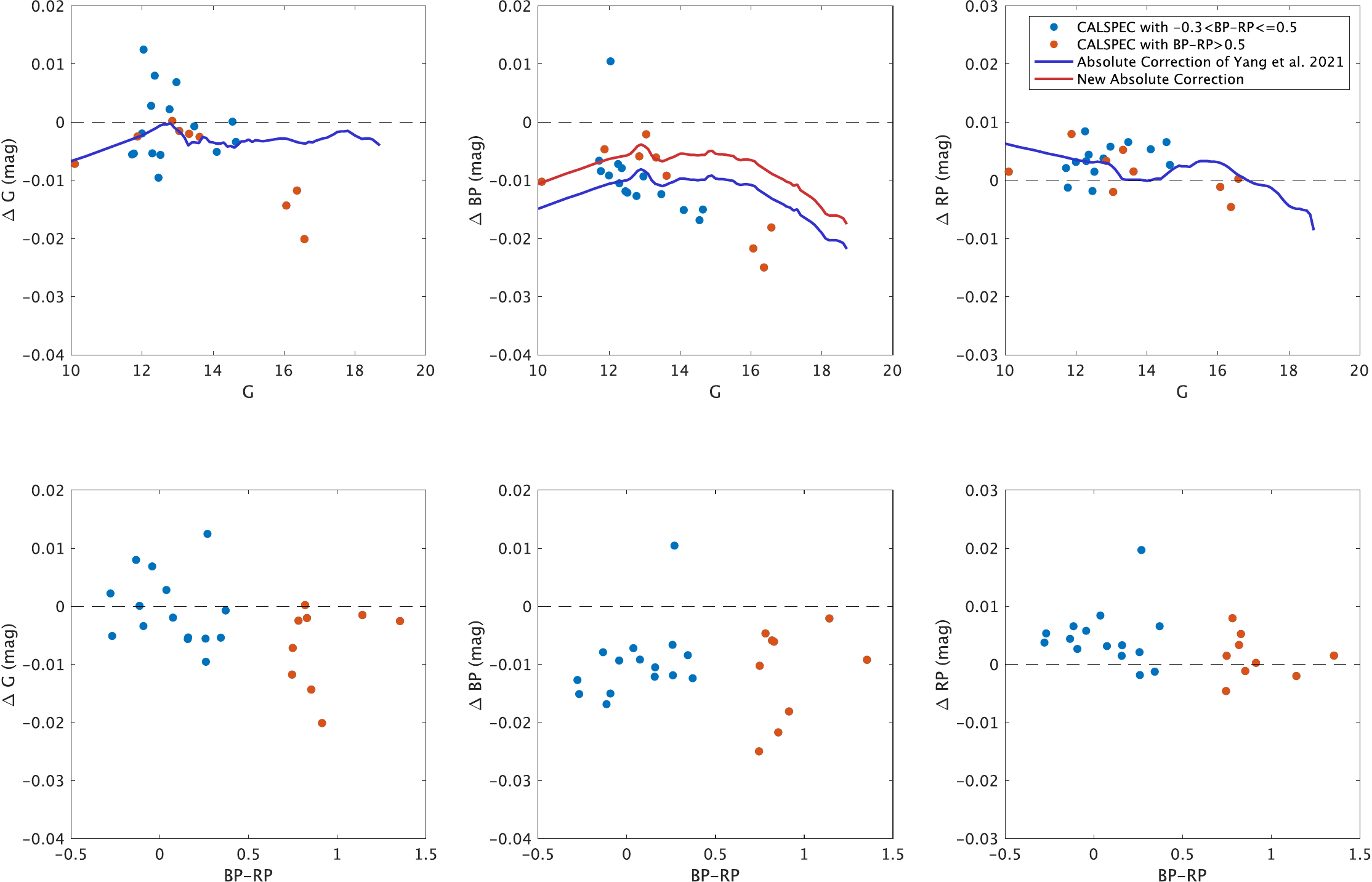}
\caption{Magnitude calibration curves, similar to the top panels of Figure\,4 in \cite{YangLinEDR3}. 
The points represent the differences between the synthetic magnitudes derived from the CALSPEC spectra and the corresponding Gaia DR3 photometry. The blue and red points are sources of $-0.3<BP-RP<0.5$ and $BP-RP>0.5$, respectively. The blue and red lines are absolute corrections of \cite{YangLinEDR3} and our new ones for the BP band, respectively.} 
\label{FigA2}
\end{figure*}

In Section\,6, we have demonstrated the color-dependent trends between the CALSPEC and the corrected Gaia DR3 photometry of \cite{YangLinEDR3} in the sixth column of Figure\,\ref{Fig16}. 
According to the color distributions of stars used in \cite{YangLinEDR3}, they employed the CALSPEC sources of much bluer colors ($-0.3<BP-RP<1.5$; more than half of CALSPEC sources have $BP-RP<0.5$) compared to the Landolt stars ($0.5 < BP-RP < 1.5$) to compute the absolute zero-point correction (see their Figure\,4). 
This resulted a zero-point offset in the absolute calibration, particularly noticeable in the $BP$ band. We therefore propose to adjust the absolute zero-point of the $BP$ band in \cite{YangLinEDR3} by 4.3 mmag, i.e., the $\Delta G_{BP}$ values in Table\,1 of \cite{YangLinEDR3} should all be added by 4.3 mmag.
This number is derived from the synthetic magnitudes of the original CALSPEC spectra of $BP-RP>0.5$ and  $10 < G <15 $, as shown in Figure\,\ref{FigA2}.

\vspace{7mm} \noindent {\bf Acknowledgments}
The authors thank the referee for the suggestions that improved the clarity of our presentation.
We acknowledge helpful discussions with Dr. Dafydd W. Evans and Dr. Francesca De Angeli. This project was developed in part at the 2023 Gaia XPloration, hosted by the Institute of Astronomy, Cambridge University.
This work is supported by the National Natural Science Foundation of China through the projects NSFC 12222301, 12173007, and 11603002, and
the National Key Basic R\&D Program of China via 2019YFA0405503. 
We acknowledge the science research grants from the China Manned Space Project with Nos. CMS-CSST-2021-A08 and CMS-CSST-2021-A09.
Z.X.N acknowledges support from the NSFC through grant No.~12303039.

Data resources are supported by China National Astronomical Data Center (NADC) and Chinese Virtual Observatory (China-VO). This work is supported by Astronomical Big Data Joint Research Center, co-founded by National Astronomical Observatories, Chinese Academy of Sciences and Alibaba Cloud.

This work has made use of data from the European Space Agency (ESA) mission {\it Gaia} (\url{https://www.cosmos.esa.int/gaia}), processed by the Gaia Data Processing and Analysis Consortium (DPAC; \url{https:// www.cosmos.esa.int/web/gaia/dpac/ consortium}). Funding for the DPAC has been provided by national institutions, in particular, the institutions participating in the Gaia Multilateral Agreement. 

Guoshoujing Telescope (the Large Sky Area Multi-Object Fiber Spectroscopic Telescope (LAMOST)) is a National Major Scientific Project built by the Chinese Academy of Sciences. Funding for the project has been provided by the National Development and Reform Commission. LAMOST is operated and managed by the National Astronomical Observatories, Chinese Academy of Sciences.

{}

\end{CJK*}

\begin{thebibliography}{}

\bibitem[Bohlin et al.(2014)]{CALSPEC14} Bohlin, R.~C., Gordon, K.~D., \& Tremblay, P.-E.\ 2014, \pasp, 126, 711. doi:10.1086/677655
\bibitem[Bohlin et al.(2019)]{Bohlin2019} Bohlin, R.~C., Deustua, S.~E., \& de Rosa, G.\ 2019, \aj, 158, 211. doi:10.3847/1538-3881/ab480c
\bibitem[Bohlin \& Lockwood(2022)]{CALSPEC22} Bohlin, R.~C. \& Lockwood, S.\ 2022, Instrument Science Report STIS 2022-7, 11 pages


\bibitem[Carrasco et al.(2021)]{dr3intcali} Carrasco, J.~M., Weiler, M., Jordi, C., et al.\ 2021, \aap, 652, A86. doi:10.1051/0004-6361/202141249
\bibitem[Cenarro et al.(2019)]{JPLUS} Cenarro, A.~J., Moles, M., Crist{\'o}bal-Hornillos, D., et al.\ 2019, \aap, 622, A176. doi:10.1051/0004-6361/201833036
\bibitem[Clem \& Landolt(2013)]{landolt2013} Clem, J.~L. \& Landolt, A.~U.\ 2013, \aj, 146, 88. doi:10.1088/0004-6256/146/4/88

\bibitem[De Angeli et al.(2022)]{GaiaXPVad} De Angeli, F., Weiler, M., Montegriffo, P., et al.\ 2022, arXiv:2206.06143. doi:10.48550/arXiv.2206.06143
\bibitem[Falc{\'o}n-Barroso et al.(2011)]{MILES2011} Falc{\'o}n-Barroso, J., S{\'a}nchez-Bl{\'a}zquez, P., Vazdekis, A., et al.\ 2011, \aap, 532, A95. doi:10.1051/0004-6361/201116842
\bibitem[Fitzpatrick(1999)]{F99} Fitzpatrick, E.~L.\ 1999, \pasp, 111, 63. doi:10.1086/316293



\bibitem[Gaia Collaboration et al.(2016)]{GaiaMission2016a} Gaia Collaboration, Prusti, T., de Bruijne, J.~H.~J., et al.\ 2016, \aap, 595, A1. doi:10.1051/0004-6361/201629272
\bibitem[Gaia Collaboration et al.(2022)]{dr3content} Gaia Collaboration, Vallenari, A., Brown, A.~G.~A., et al.\ 2022, arXiv:2208.00211. doi:10.48550/arXiv.2208.00211
\bibitem[Gaia Collaboration et al.(2022)]{GaiaDR32022} Gaia Collaboration, Vallenari, A., Brown, A.~G.~A., et al.\ 2022, arXiv:2208.00211. doi:10.48550/arXiv.2208.00211
\bibitem[Gaia Collaboration et al.(2022)]{GSPC} Gaia Collaboration, Montegriffo, P., Bellazzini, M., et al.\ 2022, arXiv:2206.06215. doi:10.48550/arXiv.2206.06215


\bibitem[Heap \& Lindler(2007)]{NGSL2007} Heap, S.~R. \& Lindler, D.~J.\ 2007, From Stars to Galaxies: Building the Pieces to Build Up the Universe, 374, 409
\bibitem[Heap \& Lindler(2009)]{NGSL2009} Heap, S. \& Lindler, D.~J.\ 2009, New Quests in Stellar Astrophysics. II. Ultraviolet Properties of Evolved Stellar Populations, 7, 273. doi:10.1007/978-0-387-87621-4\_37
\bibitem[Huang \& Yuan(2022)]{2022S82} Huang, B. \& Yuan, H.\ 2022, \apjs, 259, 26. doi:10.3847/1538-4365/ac470d
\bibitem[Koleva \& Vazdekis(2012)]{NGSL} Koleva, M. \& Vazdekis, A.\ 2012, \aap, 538, A143. doi:10.1051/0004-6361/201118065
\bibitem[Luo et al.(2015)]{LAMOST} Luo, A.-L., Zhao, Y.-H., Zhao, G., et al.\ 2015, Research in Astronomy and Astrophysics, 15, 1095. doi:10.1088/1674-4527/15/8/002


\bibitem[Mendes de Oliveira et al.(2019)]{SPLUS} Mendes de Oliveira, C., Ribeiro, T., Schoenell, W., et al.\ 2019, \mnras, 489, 241. doi:10.1093/mnras/stz1985
\bibitem[Montegriffo et al.(2022)]{dr3extcali} Montegriffo, P., De Angeli, F., Andrae, R., et al.\ 2022, arXiv:2206.06205. doi:10.48550/arXiv.2206.06205
\bibitem[Niu et al.(2021)]{Niu2021a} Niu, Z., Yuan, H., \& Liu, J.\ 2021, \apj, 909, 48. doi:10.3847/1538-4357/abdbac

\bibitem[Ruz-Mieres(2022)]{GaiaXPy} Ruz-Mieres, D.\ 2022, gaia-dpci/GaiaXPy: GaiaXPy 1.2.0, Zenodo, doi:10.5281/zenodo.7015044

\bibitem[S{\'a}nchez-Bl{\'a}zquez et al.(2006)]{MILES} S{\'a}nchez-Bl{\'a}zquez, P., Peletier, R.~F., Jim{\'e}nez-Vicente, J., et al.\ 2006, \mnras, 371, 703. doi:10.1111/j.1365-2966.2006.10699.x
\bibitem[Schlegel et al.(1998)]{SFD98} Schlegel, D.~J., Finkbeiner, D.~P., \& Davis, M.\ 1998, \apj, 500, 525. doi:10.1086/305772
\bibitem[Schlafly \& Finkbeiner(2011)]{Schlafly2011} Schlafly, E.~F. \& Finkbeiner, D.~P.\ 2011, \apj, 737, 103. doi:10.1088/0004-637X/737/2/103


\bibitem[Wang et al.(2018)]{LEMONY} Wang, C., Liu, X.-W., Huang, Y., et al.\ 2018, \mnras, 480, 4766. doi:10.1093/mnras/sty2069
\bibitem[Wu et al.(2011)]{LASP} Wu, Y., Luo, A.-L., Li, H.-N., et al.\ 2011, Research in Astronomy and Astrophysics, 11, 924. doi:10.1088/1674-4527/11/8/006


\bibitem[Xiao et al.(2023a)]{XiaoJPLUS} Xiao, K., Yuan, H., L{\'o}pez-Sanjuan, C., et al.\ 2023a, arXiv:2309.11225. doi:10.48550/arXiv.2309.11225
\bibitem[Xiao et al.(2023b)]{XiaoSPLUS} Xiao, K., Huang, Y., Yuan, H., et al.\ 2023b, arXiv:2309.11533. doi:10.48550/arXiv.2309.11533

\bibitem[Xu et al.(2022)]{Xushuai} Xu, S., Yuan, H., Niu, Z., et al.\ 2022, \apjs, 258, 44. doi:10.3847/1538-4365/ac3df6
\bibitem[Yang et al.(2021)]{YangLinEDR3} Yang, L., Yuan, H., Zhang, R., et al.\ 2021, \apjl, 908, L24. doi:10.3847/2041-8213/abdbae
\bibitem[Yuan et al.(2013)]{Yuan2013} Yuan, H.~B., Liu, X.~W., \& Xiang, M.~S.\ 2013, \mnras, 430, 2188. doi:10.1093/mnras/stt039
\bibitem[Yuan et al.(2015)]{SCR2015a} Yuan, H., Liu, X., Xiang, M., et al.\ 2015, \apj, 799, 133. doi:10.1088/0004-637X/799/2/133
\bibitem[Zhang et al.(2023)]{GaiaPara} Zhang, X., Green, G.~M., \& Rix, H.-W.\ 2023, arXiv:2303.03420. doi:10.48550/arXiv.2303.03420



\end{thebibliography}
\end{document}